\documentclass[letterpaper,prl,aps,twocolumn,showpacs,superscriptaddress]{revtex4-1}
\usepackage{bm,color}
\usepackage{mathptmx}
\usepackage[T1]{fontenc}
\usepackage[utf8]{inputenc}
\usepackage{graphicx} 
\usepackage{amsmath} 
\usepackage{amsfonts} 
\usepackage{braket}

\def \br{{\bf r}}

\def \bq{{\bf q}}
\def \bk{{\bf k}}

\def \bz{{\bf z}}
\def \bx{{\bf x}}

\def \be{{\bf e}}

\newcommand{\la}{\langle}
\newcommand{\ra}{\rangle}
\newcommand{\lla}{\langle\!\langle}
\newcommand{\rra}{\rangle\!\rangle}
\newcommand{\Ho}{\hat{H}}
\newcommand{\Uo}{\hat{U}}
\newcommand{\Go}{\hat{G}}
\renewcommand{\ao}{\hat{a}}
\renewcommand{\aa}{\hat{a}^\dag}

\newcommand{\no}{\hat{n}}

\renewcommand{\bk}{{\bm k}}
\renewcommand{\bz}{{\bm 0}}

\newcommand{\bh}{{\bm h}}
\newcommand{\bF}{{\bm F}}
\renewcommand{\br}{{\bm r}}
\renewcommand{\bq}{{\bm q}}
\renewcommand{\bx}{{\bm x}}
\newcommand{\bxi}{{\bm \xi}}
\newcommand{\rd}{{\mathrm d}}
\renewcommand{\be}{\begin{equation}}
\newcommand{\ee}{\end{equation}}
\newcommand{\bes}{\begin{eqnarray}}
\newcommand{\ees}{\end{eqnarray}}

	\newcommand{\mbf}[1]{\mathbf{#1}}
	\newcommand{\bbm}{\begin{pmatrix}}
	\newcommand{\ebm}{\end{pmatrix}}

\begin{document}

\title{Measuring topology by dynamics: Chern number from linking number}

\author{Matthias Tarnowski}
\affiliation{Institut für Laserphysik, Universität Hamburg, 22761 Hamburg, Germany}
\affiliation{The Hamburg Centre for Ultrafast Imaging, 22761 Hamburg, Germany}
\author{F. Nur Ünal}
\affiliation{Max-Planck-Institut für Physik komplexer Systeme, Nöthnitzer Straße 38, 01187 Dresden, Germany}
\author{Nick Fläschner}
\affiliation{Institut für Laserphysik, Universität Hamburg, 22761 Hamburg, Germany}
\affiliation{The Hamburg Centre for Ultrafast Imaging, 22761 Hamburg, Germany}
\author{Benno S. Rem}
\affiliation{Institut für Laserphysik, Universität Hamburg, 22761 Hamburg, Germany}
\affiliation{The Hamburg Centre for Ultrafast Imaging, 22761 Hamburg, Germany}
\author{André Eckardt}
\affiliation{Max-Planck-Institut für Physik komplexer Systeme, Nöthnitzer Straße 38, 01187 Dresden, Germany}
\author{Klaus Sengstock}
\email{klaus.sengstock@physnet.uni-hamburg.de}
\affiliation{Institut für Laserphysik, Universität Hamburg, 22761 Hamburg, Germany}
\affiliation{The Hamburg Centre for Ultrafast Imaging, 22761 Hamburg, Germany}
\affiliation{Zentrum für Optische Quantentechnologien, Universität Hamburg, 22761 Hamburg, Germany}
\author{Christof Weitenberg}
\affiliation{Institut für Laserphysik, Universität Hamburg, 22761 Hamburg, Germany}
\affiliation{The Hamburg Centre for Ultrafast Imaging, 22761 Hamburg, Germany}

\date{\today}

\begin{abstract}
Integer-valued topological indices, characterizing nonlocal properties of quantum states of matter, are known to directly predict robust physical properties of equilibrium systems. The Chern number, e.g., determines the quantized Hall conductivity of an insulator. Using fermionic atoms in a periodically driven optical lattice, here we demonstrate experimentally that the Chern number determines also the far-from-equilibrium dynamics of a quantum system. Following the proposal of ref.~[Wang et al., Phys. Rev. Lett. 118, 185701 (2017)] and extending it to Floquet systems, we measure the linking number that characterizes the trajectories of momentum-space vortices emerging after a strong quench.  We observe that it directly corresponds to the ground-state Chern number. This one-to-one relation between a dynamical and a static topological index allows us to experimentally map out the phase diagram of our system. Furthermore, we measure the instantaneous Chern number and show that it remains zero under the unitary dynamics.
\end{abstract}

\maketitle

Topological quantum matter has recently received much attention, because it opens an entirely new class of quantum phases and has potential applications ranging from precision measurements to quantum information processing and spintronics \cite{Hasan2010}. An important role is played by the Chern number, which characterizes the topology of filled bands in two-dimensional lattice systems and also underlies the integer Quantum Hall effect. Ultracold quantum gases are a promising experimental platform to explore these questions. On the one hand they allow for the realization of topologically non-trivial band structures and artificial gauge fields \cite{Lin2011,Struck2011,Struck2012,Jotzu2014,Aidelsburger2015,Kennedy2015,Flaschner2016,Tarnowski2017} and on the other hand typical time scales for dynamical studies are experimentally accessible. Moreover, they offer the perspective of combining these effects with strong interactions (see, e.g., refs. \cite{Cooper2013,Grushin2014,Anisimovas2015}). In cold atom systems, the Chern number was measured for the Hofstadter model \cite{Aidelsburger2015} using transport measurements and for the Haldane model using quantized circular dichroism \cite{Asteria2018}. 

\begin{figure}
		\includegraphics[width=0.9\linewidth]{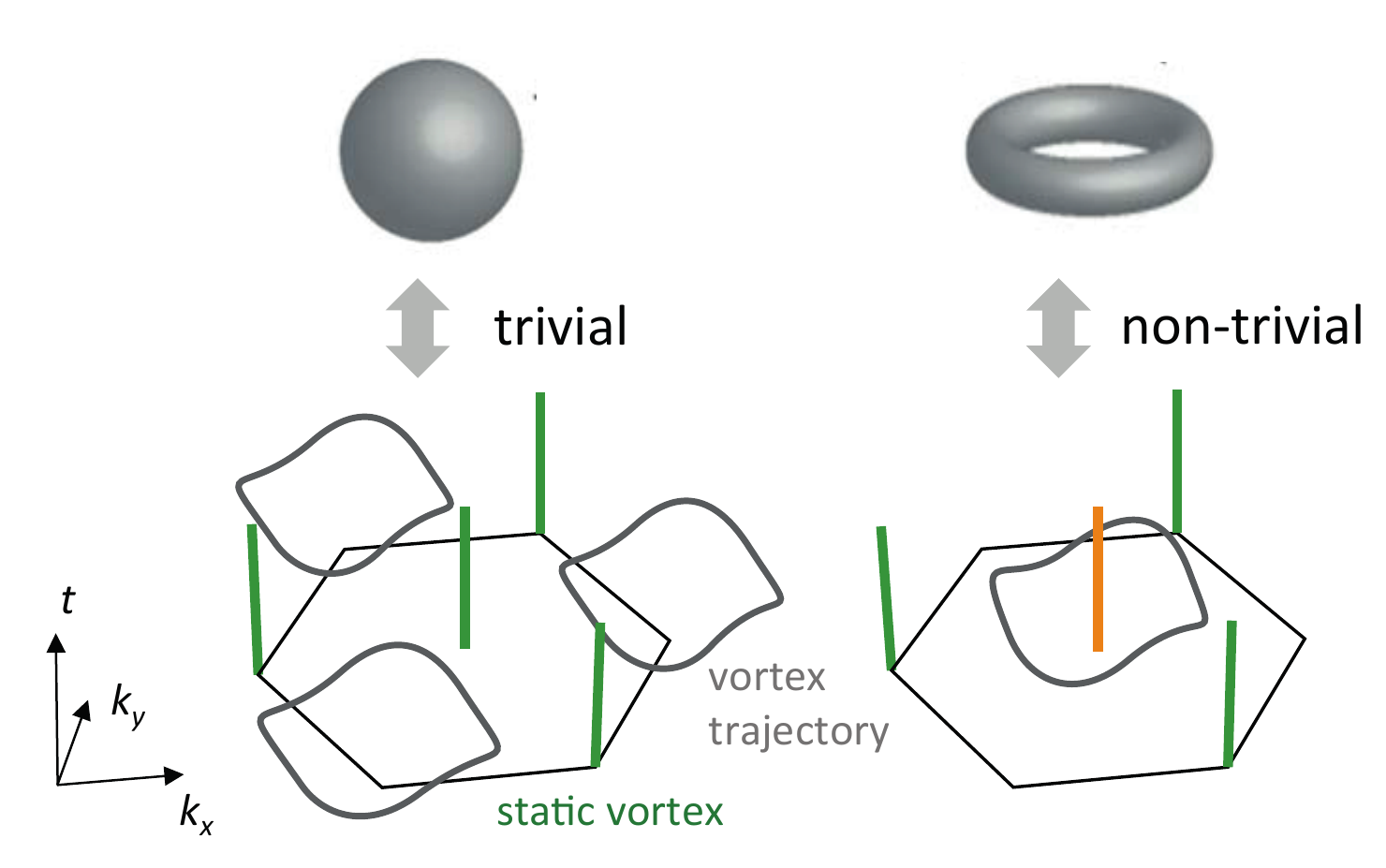}
		\caption{{\bf Illustration of the linking number of dynamical vortices}. The figure shows the Brillouin zone (hexagon) and trajectories of momentum-space phase vortices. One can define a linking number between the static vortices (straight green line) and the dynamical vortex contour (grey closed line). The linking number is zero in the left panel and one in the right panel, which can be directly mapped to the Chern number of the underlying Hamiltonian (illustrated by the sphere and torus).
}\label{fig:1_scheme}
	\end{figure}

Here we experimentally investigate a fascinating connection between the topological properties of the ground state and its far-from-equilibrium dynamics following a strong quench that was recently proposed in ref.\,\cite{Wang2017}. Using time-resolved state tomography of the time-evolved wave function, we show that the instantaneous Chern number remains zero during the dynamics. Furthermore, the state tomography reveals two kinds of vortices in momentum space: i) static vortices indicating the Dirac points and ii) dynamical vortices, which appear and disappear in pairs and trace out a closed contour \cite{Flaschner2017}. Whether this contour encloses one of the static vortices or not is a topological index (called linking number), which directly corresponds to the ground-state Chern number of the post-quench Hamiltonian \cite{Wang2017} (see Fig.\,\ref{fig:1_scheme}). Using this correspondence, we map out the phase diagram of a Floquet-engineered Haldane-type lattice model, characterized by different Chern numbers. A similar approach for a spin-orbit coupled band structure was recently demonstrated in ref.\,\cite{Sun2018}.

\begin{figure}
		\includegraphics[width=0.9\linewidth]{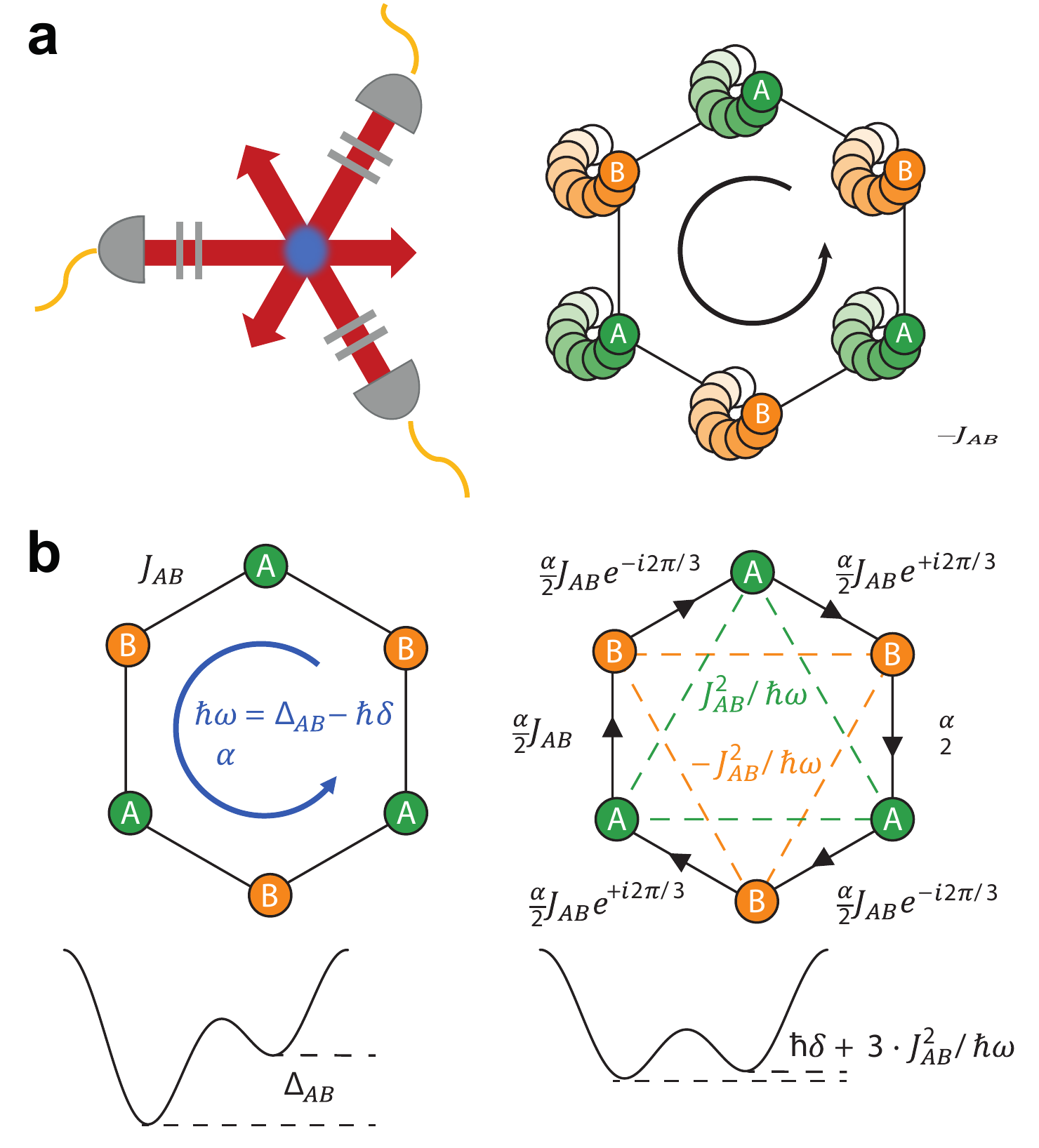}
		\caption{{\bf Experimental realization of topological bands in driven optical lattices}. {\bf a}, Three laser beams interfere under 120$^{\mathrm{o}}$ and form a hexagonal optical lattice. The geometry can be tuned via the polarisation of the lattice beams using two wave plates. The lattice can be accellerated along a circular trajectory. {\bf b}, Illustration of the tight-binding model of the bare lattice (left) and the effective Hamiltonian for the driven lattice (right). The sublattice offset in the effective Hamiltonian can be tuned via the shaking detuning $\delta$.}\label{fig:2_setup}
		\end{figure}

{\bf Floquet description of the driven hexagonal lattice.}
We start with a hexagonal optical lattice \cite{Soltan-Panahi2011} with two sublattices $A$ and $B$, which are connected by nearest-neighbor tunneling matrix elements $J_{AB}$ and have a potential offset of $\Delta_{AB}$ (see Fig.\,\ref{fig:2_setup}). It is described by the bare Hamiltonian 
\begin{equation}
\hat{H}_0=-\sum\limits_{\la l'l \ra}J_{AB}\hat{a}^{\dagger}_{l'}\hat{a}_{l}+\sum\limits_{l \in B}\Delta_{AB}\hat{n}_{l}
\end{equation} 
where $\hat{a}_{l}$ and $\hat{n}_{l}$ denote the annhiliation operator and number operator for a fermion at site $l$ and $\la l'l \ra$ denotes a pair of neighboring sites. 

\begin{figure}
		\includegraphics[width=0.9\linewidth]{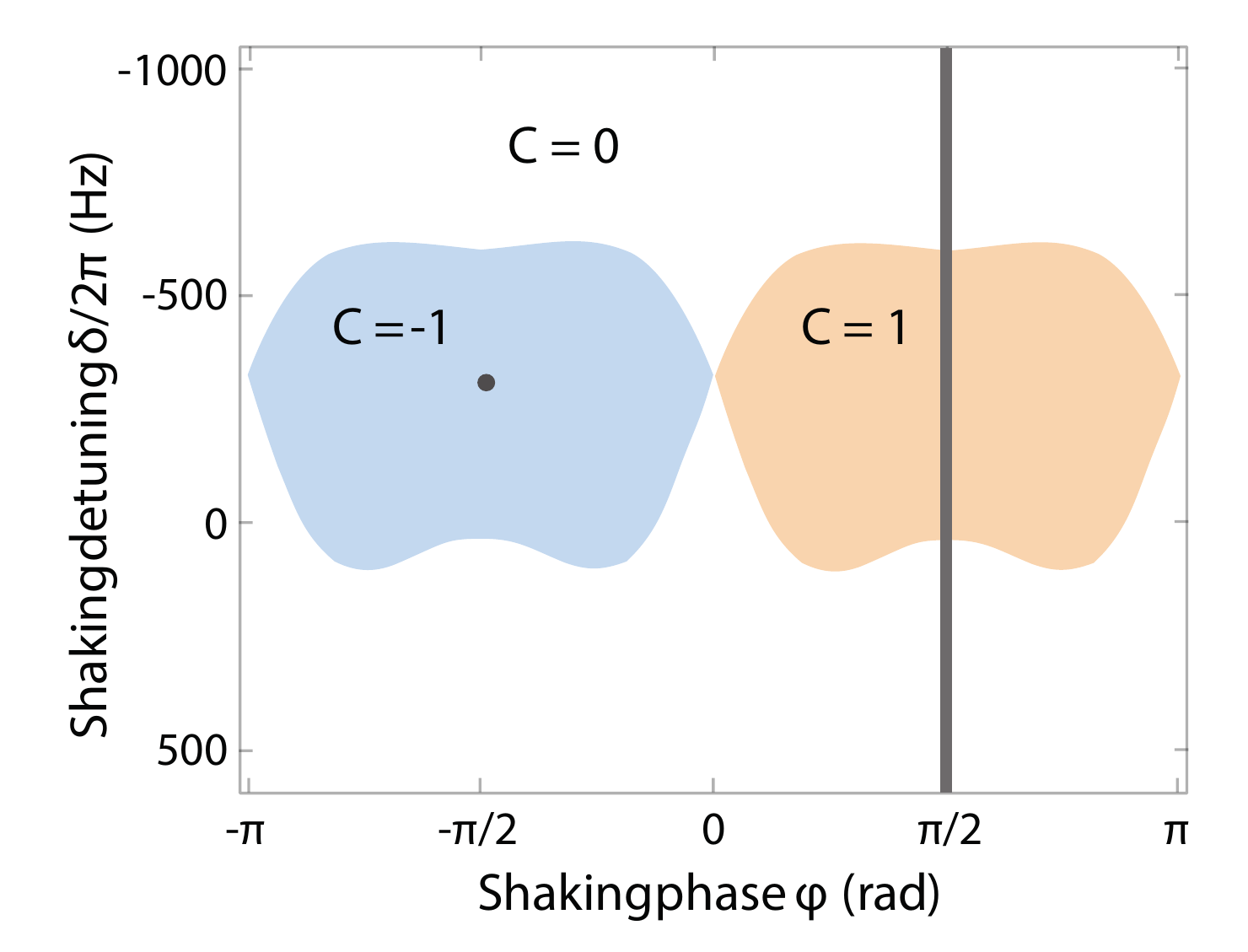}
		\caption{{\bf Topological phase diagram}. Calculated Haldane-like phase diagram of the driven hexagonal lattice with regions of different Chern number (parameters: $\omega=2\pi\cdot 6410$\,Hz, $\alpha=1.28$, the detuning $\delta$ is varied by changing the initial offset $\Delta_{AB}$ via the lattice depth). The experiments are performed for circular shaking (along the grey line and also for the grey point in the $C=-1$ area).}\label{fig:3_Haldane}
		\end{figure}

Via lattice shaking \cite{Lin2011,Struck2011,Struck2012,Jotzu2014,Aidelsburger2015,Kennedy2015,
Flaschner2016,Eckardt2005,Lignier2007,Parker2013,Goldman2014,Bukov2015,Eckardt2015,
Eckardt2017} we induce a circular inertial force of angular frequency $\omega=\Delta_{AB}/\hbar-\delta$ and amplitude $F=\alpha \hbar\omega/a$, with small detuning $\delta$, site spacing $a$, and dimensionless driving strength $\alpha$. 
The resulting Floquet system is described by a time-independent effective Hamiltonian \cite{Eckardt2015}, which is given by \cite{supmat}
\begin{equation}
\hat{H}_F=-\sum\limits_{\la l'l \ra}J_{AB}^{\rm eff}\hat{a}^{\dagger}_{l'}\hat{a}_{l}+\sum\limits_{\lla l'l \rra_A} \!\!\!J_{AA}^{\rm eff}\hat{a}^{\dagger}_{l'}\hat{a}_{l}+\sum\limits_{\lla l'l \rra_B} \!\!\!J_{BB}^{\rm eff}\hat{a}^{\dagger}_{l'}\hat{a}_{l}+\sum\limits_{l \in B}\Delta^{\rm eff}\hat{n}_{l}.
\end{equation}
In the limit of low driving strength, the expressions for the effective tunnel elements read $J_{AB}^{\rm eff}\simeq \pm \frac{\alpha}{2}J_{AB}e^{\mp i \phi_{l'l}}$ with Peierls phases $\phi_{l'l}$ for NN tunneling and $J_{AA}^{\rm eff}=-J_{BB}^{\rm eff}\simeq J_{AB}^2/\hbar\omega$ for next-nearest neighbor (NNN) tunneling, which arises as a super-exchange process. The effective sublattice offset becomes $\Delta^{\rm eff}=\hbar\delta+3J_{AB}^2/\hbar\omega$ (see Fig.\,\ref{fig:2_setup}{\bf b}). Note that in contrast to the case without initial sublattice offset \cite{Jotzu2014,Oka2009,Rechtsman2013}, we realize the Hamiltonian in a gauge, where the Peierls phases appear at the NN tunneling, which gives rise to a shifted band structure with one of the Dirac points at the $\Gamma$ point \cite{Flaschner2016}. The band structure of the Hamiltonian undergoes topological phase transitions between different lowest-band Chern numbers $C=0$ and $C=\pm 1$ at the parameters $\hbar\delta\simeq -15 J_{AB}^2/\hbar\omega$ and $\hbar\delta\simeq 3 J_{AB}^2/\hbar\omega$. We note that the width of the non-trivial region is broader than in the case without initial offset, because it does not rely on the size of the small initial next-nearest neighbor tunneling element \cite{supmat}. By going away from circular shaking to a general shaking phase $\phi$ between $x$ and $y$ direction, one obtains the phase diagram shown in Fig.\,\ref{fig:3_Haldane} resembling that of the Haldane model \cite{Haldane1988}. 

In this two-band model, the Hamiltonian and the time-evolved modes can be visualized on a Bloch sphere for each quasi-momentum. The Hamiltonian is diagonal with respect to quasimomentum $\mbf{k}$ and can be written in the form 
\begin{equation}
h(\mbf{k})=h_0(\mbf{k})+\mbf{\sigma}\cdot\mbf{h}(\mbf{k}).
\end{equation} 
Here $\mbf{h}(\mbf{k})$  plays the role of a "magnetic field" coupling via the vector of Pauli matrices $\mbf{\sigma}$ to the pseudo-spin-1/2 degree of freedom,  
which is spanned by the two sublattice states and represented by a unit vector ${\bf \psi}({\bf k})$ on the Bloch sphere.
It induces a  $\mbf{k}$-dependent precession dynamics with angular velocity $2|\mbf{h}(\mbf{k})|/\hbar$ around itself. Its direction $\hat{\mbf{h}}(\mbf{k})=\mbf{h}(\mbf{k})/|\mbf{h}(\mbf{k})|$ determines the two eigenstates ${\bf \psi}_\pm({\bf k})=\mp \hat{\bf h}({\bf k})$ forming both Bloch bands and, therefore, completely characterizes the topology of the system. In fact, in such a two-band system, the Chern number has a simple geometric interpretation: it counts the number of times  $\hat{\mbf{h}}(\mbf{k})$ covers the Bloch sphere for $\mbf{k}$ in the first Brillouin zone \cite{Hasan2010}. 
From the direction $\hat{\mbf{h}}(\mbf{k})$ one can obtain the Berry curvature of the lowest band
\begin{equation}
\Omega(\mbf{k})=\frac{1}{2}(\partial k_x \hat{\mbf{h}}(\mbf{k}) \times \partial k_y \hat{\mbf{h}}(\mbf{k}))\cdot \hat{\mbf{h}}(\mbf{k}).\label{eq_BC}
\end{equation}
and the corresponding Chern number $C=\frac{1}{2\pi}\int d^2k \Omega(\mbf{k})$ by integration over the first Brillouin zone.

\begin{figure}[tbp]
	\centering
		\includegraphics[width=0.9\linewidth]{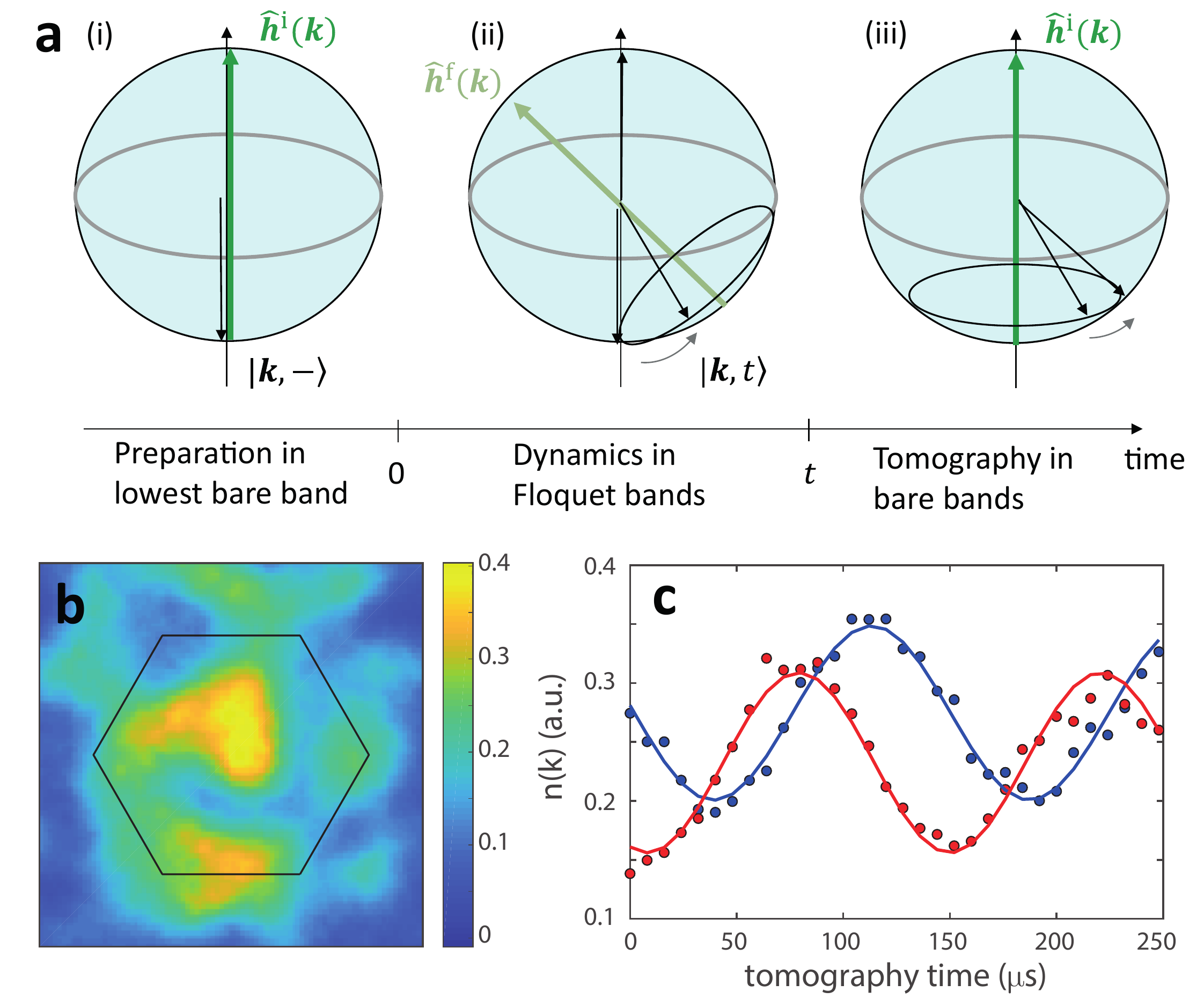}
		\caption{{\bf Illustration of the time-resolved state tomography.} {\bf a}, The states of the two-band model can be visualized on a Bloch sphere with the eigenstates of the two bare bands on the poles. (i) We initialize the state of the system in the lower bare band of $\hat{\mbf{h}}^{\rm i}(\mbf{k})$  (south pole). (ii) We quench into the final Floquet system $\hat{\mbf{h}}^{\rm f}(\mbf{k})$  by suddenly switching on the lattice shaking. The states (black arrow) evolve on the Bloch sphere according to the Floquet Hamiltonian. (iii) We measure the time-evolved state by projecting back onto the bare bands of $\hat{\mbf{h}}^{\rm i}(\mbf{k})$  and following the dynamics. When the time evolved state $\ket{\mbf{k},t}$ is at one of the poles, this leads to the absence of dynamics in the tomography and to a vortex in the azimuthal phase profile. 
{\bf b}, Example image of the momentum density $n(\mbf{k})$ obtained by time-of-flight expansion for detuning $\delta=-2\pi\cdot 372$\,Hz, evolution time $t = 0.429$\,ms and tomography time $t' = 104\,\mu$s. The hexagon marks the first Brillouin zone.
{\bf c}, The interference of the $A$ and $B$ sublattices maps the precession onto an oscillation in the density, from which one obtains the phase $\phi(\mbf{k},t)$ and the amplitude $\sin(\theta(\mbf{k},t))$ (compare Eq.~(\ref{eq n_k(t)})). The plot shows the oscillation with the respective fit for a selected pixel in the image in {\bf b}, i.e. a single momentum state, and for the evolution time $t=0.429$\,ms (blue) and $t=0.624$\,ms (red) in the Floquet system.}\label{fig:4_scheme}
		\end{figure}

{\bf Time-resolved state tomography}
In order to access the topology of our system, we use a state tomography scheme, which was introduced in ref.~\cite{Hauke2014} and demonstrated in ref. \cite{Flaschner2016}. Here, we are interested in the dynamics of the state after a quench between two Hamiltonians $\mbf{h}^{\rm i}(\mbf{k})$ and $\mbf{h}^{\rm f}(\mbf{k})$ and use a time-resolved state tomography \cite{Flaschner2017}, which involves a projection onto a tomography Hamiltonian $\mbf{h}^{\rm t}(\mbf{k})$, i.e. a double quench protocol, as illustrated in Fig.\,\ref{fig:4_scheme} for the special case $\mbf{h}^{\rm t}(\mbf{k})=\mbf{h}^{\rm i}(\mbf{k})$. 

The key idea of the tomography is to observe a precession under the action of the tomography Hamiltonian, which can be observed in time-of-flight measurements.
The momentum distribution after a time-of-flight measurement on a state given by
$|\psi(\mbf{k})\rangle= \cos(\theta(\mbf{k})/2) |A\rangle + \sin(\theta(\mbf{k})/2) e^{i\phi(\mbf{k})}|B\rangle$, can be expressed as
\begin{equation}\label{eq n_k(t)}
\begin{split}
n(\mbf{k})&=f(\mbf{k})|\langle A|\psi(\mbf{k})\rangle + \langle B|\psi(\mbf{k})\rangle |^2 \\
&=f(\mbf{k}) \{1+\sin(\theta(\mbf{k}))\cos(\phi(\mbf{k})) \},
\end{split}
\end{equation}
where$(|A\rangle, |B\rangle)$ are the poles of the Bloch sphere and $f(\mbf{k})$ is the Fourier transform of the Wannier function. This measurement is nothing but a projection onto the $\mbf{x}$-axis of the Bloch sphere, $|x\rangle=(|A\rangle+|B\rangle)/2$. One can see this easily by expressing the Bloch vector $\mbf{\psi}(\mbf{k})$ representing the wave function $|\psi(\mbf{k})\ra$ as
\begin{equation}\label{eq w.f.}
\mbf{\psi}(\mbf{k}) = 
\left( \begin{array}{c} \sin(\theta({\mbf{k}}))\cos(\phi({\mbf{k}})) \\ \sin(\theta({\mbf{k}}))\sin(\phi({\mbf{k}})) \\ \cos(\theta({\mbf{k}})) \end{array} \right)
\end{equation}
We can immediately see that the $x$-component is given by $\sin(\theta({\mbf{k}}))\cos(\phi({\mbf{k}}))$.

In all experiments described in the manuscript, we start with a filled lowest band $|\psi^{\rm i}(\mbf{k})\rangle$
of the initial Hamiltonian describing the bare lattice and quench into the final Hamiltonian describing the shaken lattice, i.e. between the two "magnetic fields" $\mbf{h}^{\rm i}(\mbf{k})$ and $\mbf{h}^{\rm f}(\mbf{k})$. After a variable evolution time, we perform state tomography in the basis of the initial lattice
by quenching to the tomography Hamiltonian $\mbf{h}^{\rm t}(\mbf{k})=\mbf{h}^{\rm i}(\mbf{k})$ and letting the system evolve for a time $t'$. The quenched state precesses around $\mbf{h}^{\rm t}(\mbf{k})$, giving rise to an oscillatory signal for the momentum distribution (\ref{eq n_k(t)}) as demonstrated in Fig.\,\ref{fig:4_scheme}(c). 

The original tomography scheme \cite{Hauke2014,Flaschner2016,Flaschner2017} and the proposal for the linking number ref.\,\cite{Wang2017} assume that the tomography Hamiltonian $\mbf{h}^{\rm t}(\mbf{k})$ is diagonal in the sublattice-basis, i.e. corresponds to completely decoupled $A$ and $B$ sublattices and with flat dispersion relations. In that case, one directly measures the angles $\theta({\mbf{k}})$ and $\phi({\mbf{k}})$ defined above and can straight-forwardly obtain the Berry curvature of the lowest band via Eq.~(\ref{eq_BC}).

{\bf State tomography with dispersive bands.}
Here, we extend this concept to a state tomography in dispersive bands, i.e. $\mbf{h}^{\rm t}(\mbf{k})$ being non-diagonal in the sublattice-basis and $\mbf{k}$-dependent.
Because both the initial and the tomography Hamiltonian are realized as the same static lattice, this allows us to start with dispersive bands ($J_{AB}/\Delta_{AB}\simeq 0.08$), yielding a much broader non-trivial region ($18 J_{AB}^2/\hbar\omega\simeq h \cdot 500$\,Hz, see above), which is easier to access experimentally even in the presence of an external trap. As a central result, we find that the topological properties can be faithfully obtained from the tomography in dispersive bands, as long as the tomography basis is itself topologically trivial, which is always ensured when using the static optical lattice for the tomography. This also demonstrates the topological robustness in our system against distortions. We note that a measurement in the diagonal basis, i.e. corresponding to completely flat bands, is possible via Stern-Gerlach separation when using internal atomic levels as spin instead of the sublattice pseudospin to generate topological structures \cite{Alba2011,Sun2018}.

While the phase profile is, in general distorted for tomography in non-flat bands, we will show here that the topological information encoded in the vortex trajectories is not altered. 
Since the linking number of the vortex trajectories can only have discrete quantized values, it is topologically protected and cannot be changed by the distorted phase profile measured in the tomography in dispersive bands. This robustness of topological defects is a general feature and was also used in the related work of ref.\,\cite{Tarnowski2017}. While the effect of dispersive bands of the initial Hamiltonian was discussed in \cite{Wang2017,Yu2017}, the effect of dispersive bands of the tomography Hamiltonian was not discussed previously.

	\begin{figure*}
		\includegraphics[width=0.7\textwidth]{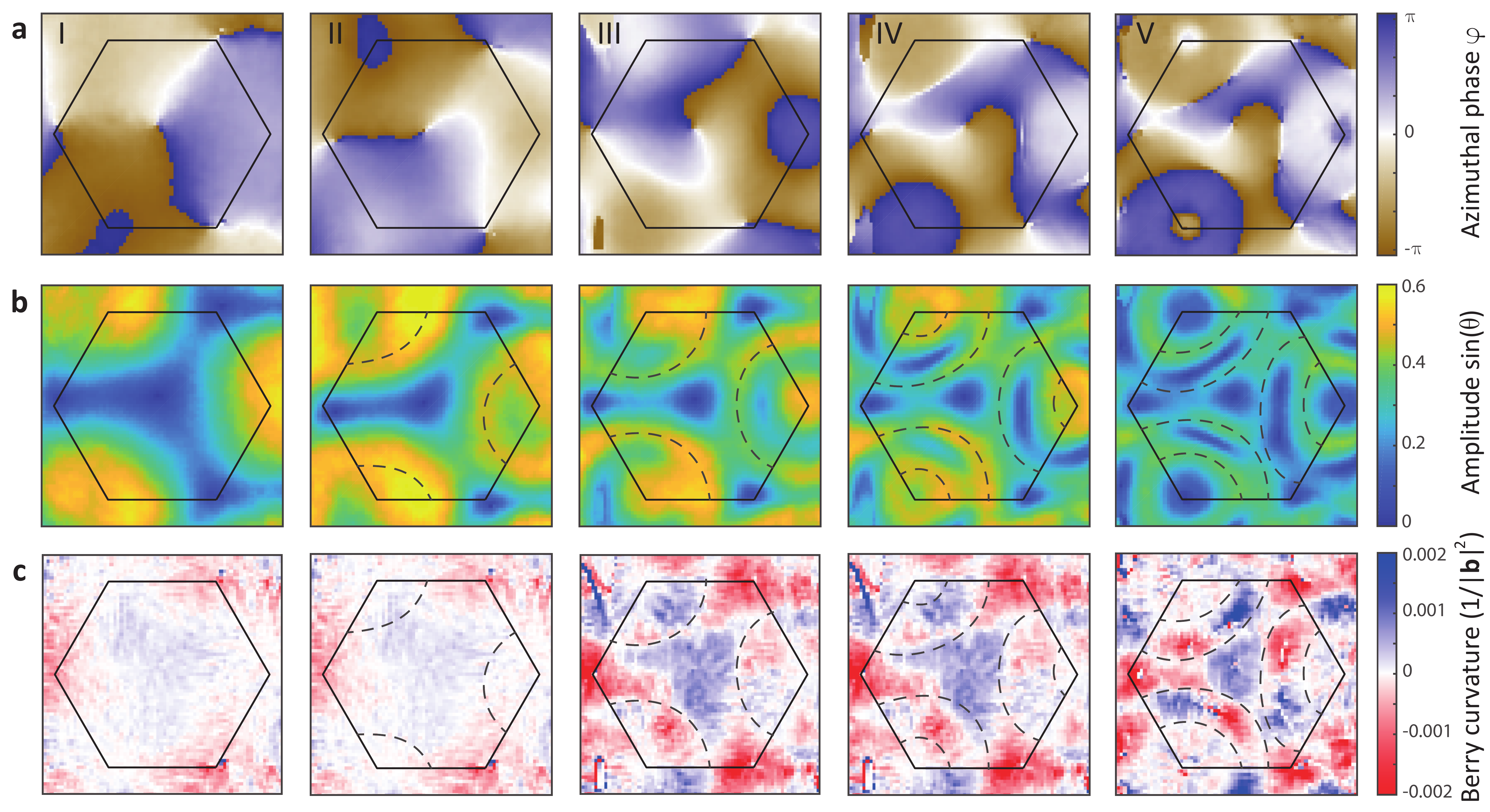}
		\caption{{\bf Instantaneous Berry curvature and instantaneous Chern number.}. Data from time-resolved state tomography in the dispersive bands showing the azimuthal phase $\phi$ ({\bf a}) and the amplitude $\sin(\theta)$ ({\bf b}) for different evolution times after the quench into a non-trivial system ($\delta=-2\pi\cdot 372$\,Hz). From these data, we obtain the instantaneous distorted Berry curvature ({\bf c}) and the instantaneous Chern number of the time-evolved state in units of the inverse squared length of a reciprocal lattice vector ${\bf b}$. The evolution times are 156, 273, 390, 507 and 663 $\mu$s and the Chern numbers are -0.001, -0.008, -0.013, -0.016, -0.015. Although the distorted Berry curvature develops finer structures as a function of time, the instantaneous Chern number stays zero, as enforced by the unitary evolution. The dashed lines separate regions, where the state lies on the southern and northern hemisphere (sketches). 
		}\label{fig:5_Chern}
	\end{figure*}

As long as the initial state is trivial, the linking numbers observed after the double-quench protocol reflects the Chern number of the ground state of the final Hamiltonian $\mbf{h}^{\rm f}(\mbf{k})$. 
Namely, a trivial dispersive band structure corresponds to a map $\mbf{h}^{\rm t}(\mbf{k})$ that does not wrap around the Bloch sphere, but covers only a part of the Bloch sphere.  
Thus, we can continuously deform $\mbf{h}^{\rm t}(\mbf{k})$ so that it points to the north pole for every $\mbf{k}$-point which is again a topologically trivial state. We do this by the $\mbf{k}$-dependent rotation defined by
\begin{equation}\label{eq rotation}
\begin{split}
\mbf{\tilde{h}}^{\rm t}(\mbf{k})&= \mbf{R}(\hat{n},\theta^{\rm t}({\mbf{k}})) \: \mbf{h}^{\rm t}(\mbf{k})\\ 
&= \mbf{R}(\sin\phi^{\rm t}({\mbf{k}}) \hat{x} - \cos\phi^{\rm t}({\mbf{k}}) \hat{y},\:\theta^{\rm t}({\mbf{k}})) \: \mbf{h}^{\rm t}(\mbf{k}),
\end{split}
\end{equation}
where the transformation is captured by the vector $\hat{\theta}^{\rm t}({\mbf{k}})$ (pointing along $\hat{n}(\mbf{k})$) and having length of $\theta^{\rm t}({\mbf{k}})$. Since $\mbf{h}^{\rm t}(\mbf{k})$ is smooth in $\mbf{k}$, fully gapped and topologically trivial, $\mbf{R}(\hat{n},\theta^{\rm t}({\mbf{k}}))$ is continuous in $\mbf{k}$ as well. When this rotation is applied to the Hamiltonians in the remaining stages of the experiment, $\mbf{h}^{\rm f}\mbf{(k)}$ and $\mbf{h}^{\rm i}\mbf{(k)}$, it does not change the topology of their band structures. Namely, it does not change the number of times $\mbf{h}^{\rm f}\mbf{(k)}$ wraps around the Bloch sphere. One can consider a patch in the $\mbf{k}$-space and its image under the map $\mbf{h}^{\rm f}(\mbf{k})$. The continuous rotation $\mbf{R}(\hat{n},\theta^{\rm t}({\mbf{k}}))$ can stretch, compress, rotate, or shift this patch on the sphere but can not cut it open. Once we perform this rotation on the Hamiltonians in all three stages, $\mbf{h}^{\rm i}(\mbf{k}),\mbf{h}^{\rm f}(\mbf{k}) $ and $\mbf{h}^{\rm t}(\mbf{k})$ (assuming $\mbf{h}^{\rm t}(\mbf{k})=\mbf{h}^{\rm i}(\mbf{k})$), the rest of the discussion follows as described by Wang et al. in ref.~\cite{Wang2017}.

In this rotated frame, the tomography Hamiltonian $\mbf{\tilde{h}}^{\rm t}(\mbf{k})$ is parallel to the $z$-axis and again we have a precession around the $z$-axis. But now the signal that we measure is the projection on the rotated $x$-axis. As a result, the phase of the measured oscillatory dynamics is not the azimuthal angle of $|\psi^{\rm f}(\mbf{k},t)\rangle=e^{-i\mbf{h}^{\rm f}(\mbf{k})\cdot\mbf{\sigma}t} |\psi^{\rm i}(\mbf{k})\rangle$. Nevertheless, the phase distribution possesses vortices whenever $|\psi^{\rm f}(\mbf{k},t)\rangle \parallel \hat{z}$. At $\mbf{k}$-values for which $|\psi^{\rm i}(\mbf{k})\rangle \parallel \mbf{\tilde{h}}^{\rm f}(\mbf{k})$, the initial state can not precess and when projected onto $\mbf{\tilde{h}}^{\rm t}(\mbf{k})$, we observe a static singularity in the tomography. On the other hand, at some $\mbf{k}$-value, if the rotated quench Hamiltonian is perpendicular to the initial state $|\psi^{\rm i}(\mbf{k})\rangle \perp \mbf{\tilde{h}}^{\rm f}(\mbf{k})$, after some precession time $t$, the state reaches $-\mbf{\tilde{h}}^{\rm t}(\mbf{k})$ direction (effective north pole) and gives rise to a dynamic vortex in the tomography. 

{\bf Chern number from tomography in dispersive bands.}
The state tomography in dispersive, but topologically trivial bands also gives access to the Chern number. It corresponds to a reconstruction of the state in a basis, which is itself non-diagonal in the sublattice-basis and has itself finite Berry curvature. Therefore the relation of the measured angles $\theta({\mbf{k}})$ and $\phi({\mbf{k}})$ to the Berry curvature is more involved and would in principle require the knowlegde of the dispersive bands, i.e. the rotation matrix $\mbf{R}$. Instead we introduce the distorted Berry curvature by inserting $\theta({\mbf{k}})$ and $\phi({\mbf{k}})$ directly in Eq.~(\ref{eq_BC}). The integral of the distorted Berry curvature over the full Brillouin zone is quantized, just like the integral over the real Berry curvature, and is equal to the Chern number. This is the case, because the rotation matrix $\mbf{R}$ quantifying the distortion due to $\mbf{h}^{\rm t}(\mbf{k})$ does not create a monopole as long as the tomography basis is topologically trivial. More precisely, the quantization of the Chern number holds for any pseudospin texture and the rotation does not change the topology, which is still determined by the direction of the pseudospin at the Dirac points, where the rotation matrix is identity. This faithful measurement of topological proporties even in the basis of dispersive bands underlines the  versatility of the state tomography approach.

{\bf Measurement of the instantaneous Chern number.}
As a central result, we measure the instantaneous Chern number of the time-evolved state after a quench into a non-trivial Hamiltonian. We obtain the instantaneous distorted Berry curvature from the time-resolved state tomography in dispersive bands shown in Fig.\,\ref{fig:5_Chern}. We find that after the quench the state develops a strong Berry curvature with finer and finer structure, but the extracted Chern number stays very close to zero with $|C|<0.02$. This confirms the finding that the Chern number, which is dictated by the trivial Hamiltonian before the quench, cannot change under unitary dynamics \cite{DAlessio2015,Caio2015,Hu2016,Unal2016}. Recently it was suggested that, conversely, the Chern-Simons invariant in one-dimensional systems can change during dynamics \cite{McGinley2018}.

The tomography scheme cannot differentiate between the northern and southern hemisphere of the Bloch sphere, because it gives access to $\sin(\theta(\mbf{k}))$ instead of $\theta(\mbf{k})$ itself. This could in principle be complemented by adiabatic band mapping measurements \cite{Hauke2014}. Alternatively, one can identify the momenta, where the state points to the equator and changes between two hemispheres, and correct the sign of the Berry curvature correspondingly \cite{Hauke2014}. In Fig.\,\ref{fig:5_Chern}, we identify these momenta and mark them by dashed lines. These momenta are identified via a local maximum of $\sin(\theta(\mbf{k}))$, although due to damping in the system, $\sin(\theta(\mbf{k}))$ does not reach one. In the data of the distorted Berry curvature, it is evident, that the curvature cancels to zero separately for each region separated by the dashed lines. Therefore a sign correction is not necessary.

\begin{figure}
		\includegraphics[width=\linewidth]{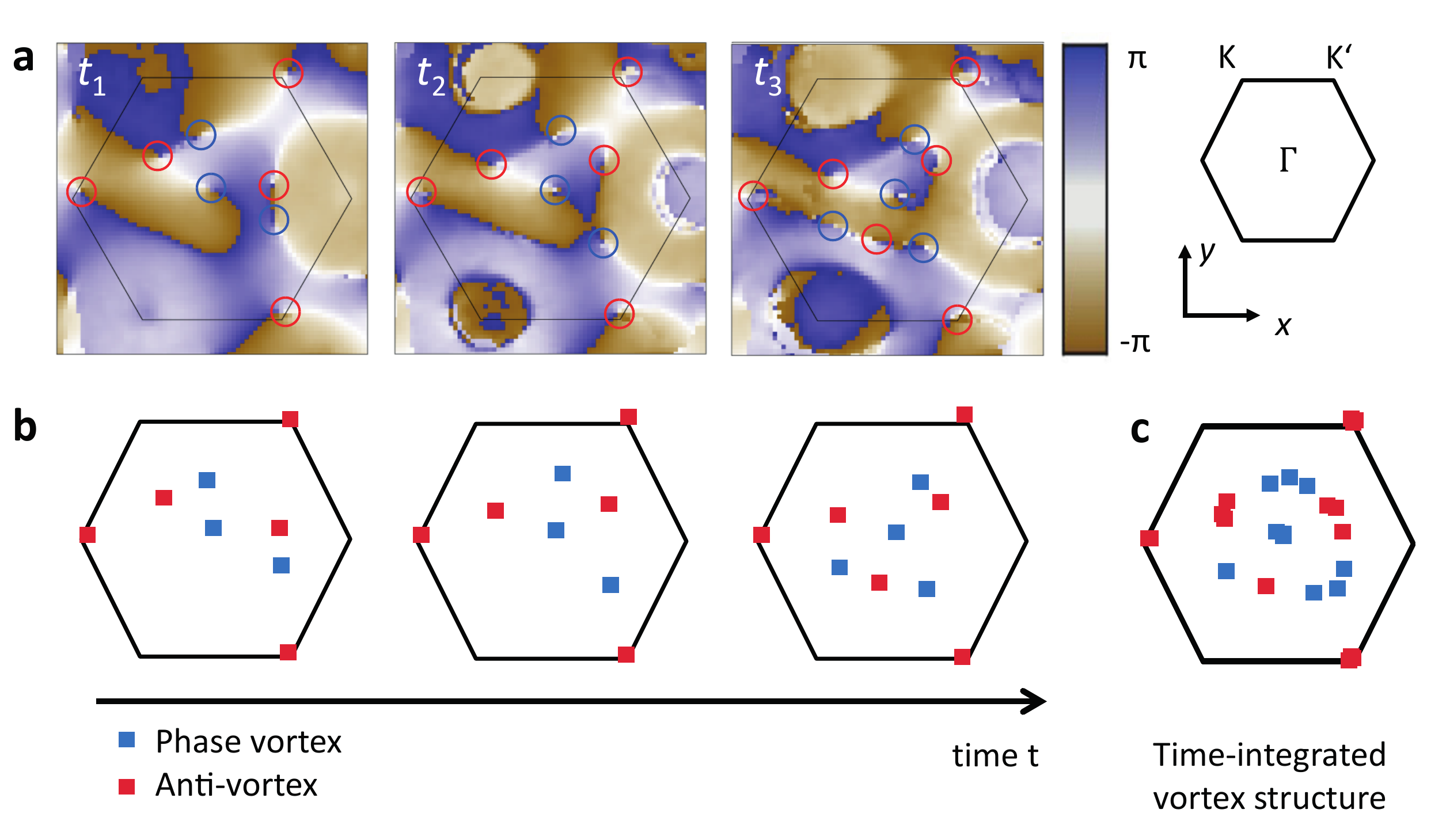}
		\caption{{\bf Extracting the vorticity of the phase profiles}. {\bf a} Azimuthal phase profile $\phi({\bf k})$ of the time-evolved state with the identified vortices marked by red and blue circles as a guide to the eye. {\bf b} Vorticity of the phase profiles with the position of the vortices and anti-vortices marked by blue and red squares, respectively. While the phase profile itself is distorted for state tomography in dispersive bands, the vortices can be clearly identified and their interpretation is not compromised. {\bf c} The time-integrated vorticity clearly shows the static vortices at the $\Gamma$ and $K'$ points and the closed contour of dynamical vortices. The detuning is $\delta=-2\pi\cdot 372$\,Hz. 
	}\label{fig:6_Vortex}
	\end{figure}

{\bf Observation of dynamical vortices}
While the time-evolved state has an instantaneous Chern number of zero independent of the Chern number of the underlying post-quench Hamiltonian, its dynamics contains information about the topology of the latter via the vortex structure of the phase profile. 
In the remainder of the manuscript, we therefore focus on the vorticity of the phase profiles of the time-resolved state tomography (see Fig.\,\ref{fig:6_Vortex}). We calculate the vorticity as $v({\bf k})=\nabla_{\bf k} \times \nabla_{\bf k} \phi({\bf k})$ and integrated it over different evolution times in the Floquet system. From this anaylsis, we clearly identify the static vortices at the $\Gamma$ and $K'$ points and the dynamical vortices, which appear and disappear in pairs and trace out a closed contour \cite{Flaschner2017} (see Fig.\,\ref{fig:6_Vortex}{\bf c}). 

\begin{figure}
		\includegraphics[width=\linewidth]{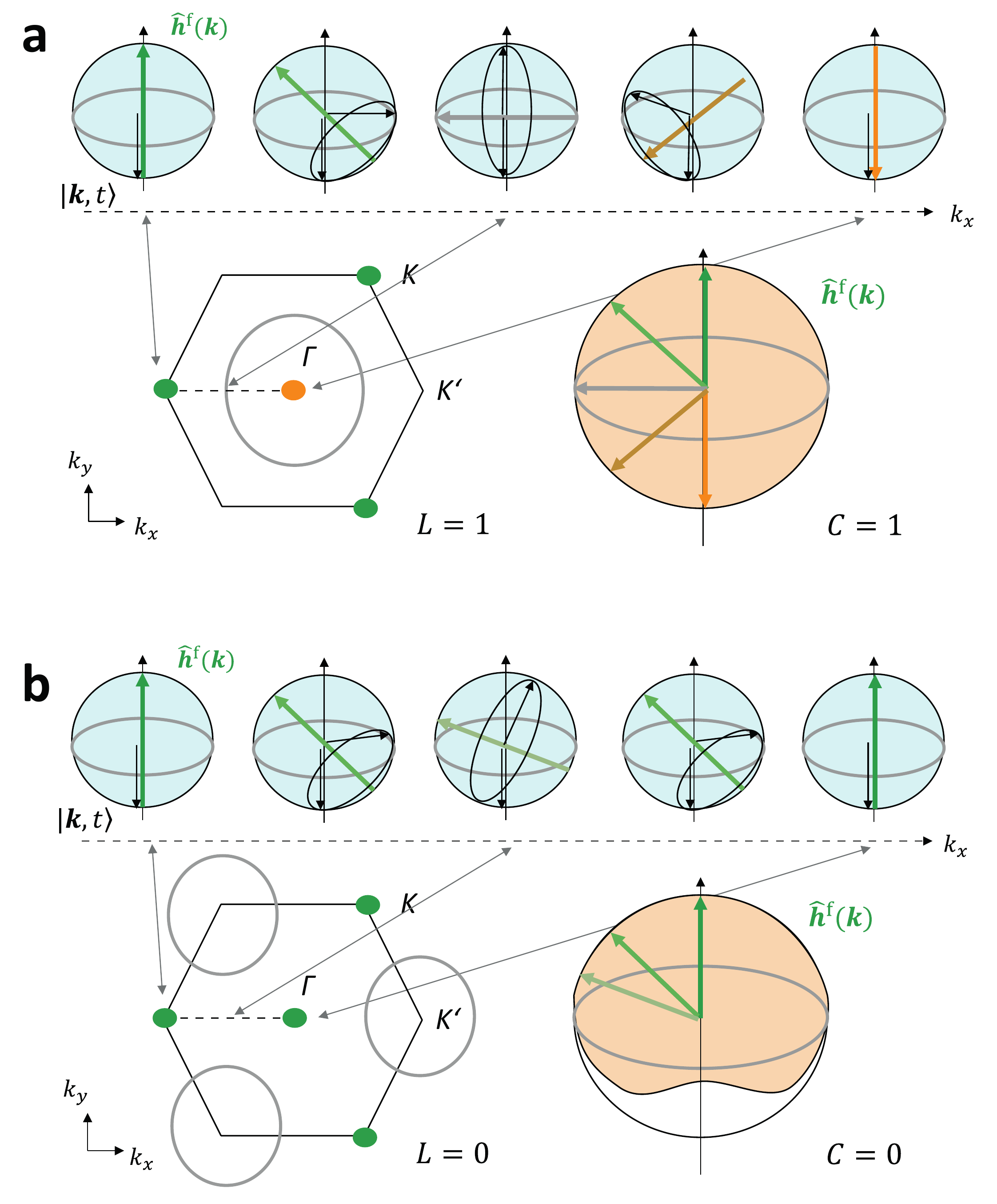}
		\caption{{\bf Illustration of the mapping between linking number and Chern number}. {\bf a}, The inverse images of the poles on the Bloch sphere form contours in the Brillouin zone. At the Dirac points, where the Hamiltonian points to one of the poles, there is no dynamics and the state will stay at the south pole and give rise to a static vortex (green and orange dots). Where the Hamiltonian lies on the equator, the time-evolved state will reach the north pole dynamically and will give rise to a dynamic vortex. These dynamic vortices move on a contour, which is the inverse image of the equator of the Bloch sphere (grey line). The Chern number of the Hamiltonian can be inferred from the linking number of the contour: if the dynamic vortex contour encloses one of the static vortices, then both static vortices correspond to opposite poles so that the Hamiltonian is topologically nontrivial. This can be seen by following the dynamics along a path connecting the two Dirac points (dashed line). In the depicted case, the Chern number is 1. {\bf b}, Same as {\bf a}, but for the case of a Chern number 0.
	}\label{fig:7_Mapping}
	\end{figure}

{\bf Mapping between Chern number and linking number of dynamical vortex contours}
As we show in the following, the Chern number of the underlying Hamiltonian maps to the linking number of these dynamical vortices, which counts whether this contour encloses one of the static vortices or not \cite{Wang2017}. The central idea is that the Chern number corresponds to the covering of the Bloch sphere, which can be measured by observing whether $\hat{\mbf{h}}(\mbf{k})$ contains both poles (see Fig.\,\ref{fig:7_Mapping}). The topology of $\hat{\mbf{h}}^{\rm f}(\mbf{k})$ is entirely encoded in the vortices of this phase profile. Namely, the linking number associated with the trajectories of vortices directly corresponds to the Chern number \cite{Wang2017,Yu2017}. While static vortices appear at the Dirac points, where $\hat{\mbf{h}}^{\rm i}(\mbf{k})$ points to one of the poles of the Bloch sphere, the contours of dynamical vortices correspond to those $\mbf{k}$ where $\hat{\mbf{h}}^{\rm i}(\mbf{k})$ points to the equator. A topologically nontrivial Hamiltonian containing both poles requires this contour to encircle a static vortex so that it has to be crossed once (or an odd number of times) between the two static vortices. The absence of a dynamical vortex contour can therefore be identified with a Chern number zero. Importantly, the topology is not signaled by the mere existence of a contour, but by its topological index: trivial contours that do not enclose a static vortex are explicitly possible.

Note that the argument can be formulated in a more general framework by considering the inverse images of any two orthogonal vectors on the Bloch sphere \cite{Wang2017}. The Chern number then maps onto the linking number of the two trajectories in the space spanned by $k_x$, $k_y$ and time and can be related to a Hopf invariant. Such a linking number characterizing a Hopf insulator was recently observed in a quantum simulation using a nitrogen vacancy center \cite{Yuan2017}.

\begin{figure*}
		\includegraphics[width=0.8\textwidth]{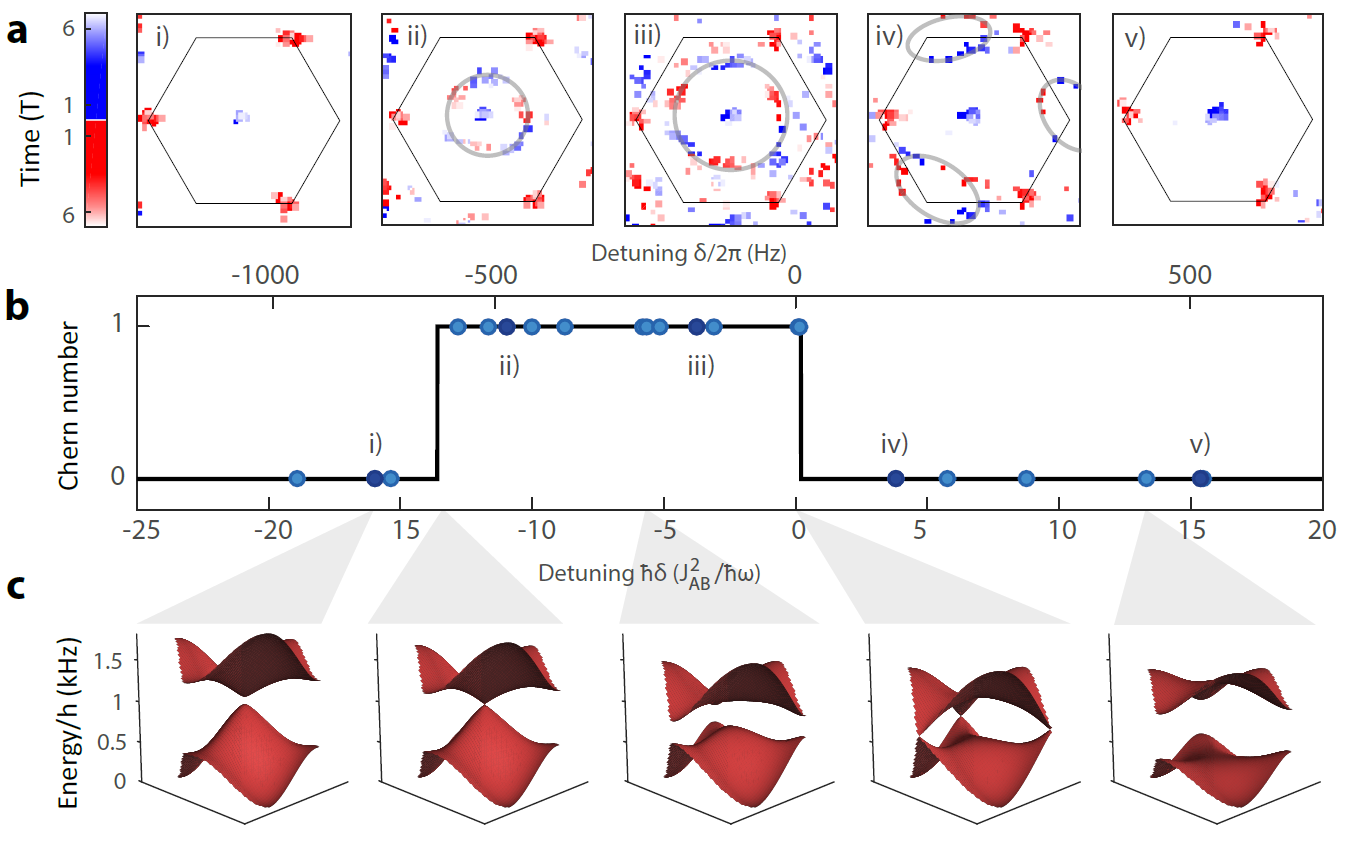}
		\caption{{\bf Mapping out the topological phase diagram using the linking number}. {\bf a}, Original data of the observed vortices summed over all time steps (red dot: positive chirality, blue dot: negative chirality; the hue indicates the time step where the vortex was present). The hexagon marks the first Brillouin zone. The dynamical vortex contours are highlighted by a guide-to-the-eye (grey line). {\bf b}, The Chern number is obtained from the linking number of these dynamical vortex contours (or the absence of a contour) and plotted for various shaking detunings (cut through the phase diagram corresponding to the grey line in Fig.\,\ref{fig:3_Haldane}). The region with non-trivial Chern number agrees well with the prediction from a full numerical calculation (solid line). {\bf c}, Calculated Floquet bands for various detunings illustrating the closing of the Dirac points at the topological phase transitions.}\label{fig:8_PhaseDiagram}
	\end{figure*}

{\bf Measurement of the topological phase diagram.}
As a central result, we use the relation between the linking number and the Chern number to experimentally map out the topological phase transition of the effective Hamiltonian. 
Fig.\,\ref{fig:8_PhaseDiagram}{\bf a} shows data of the time-integrated vorticity for different quenches into Chern $0$ and Chern $1$ areas of the phase diagram (different detunings of the lattice shaking). 
While the static vortices at the $\Gamma$ and $K$ points are visible in all data sets, one clearly recognizes additional vortex contours in the data sets for near-resonant shaking. We easily count the linking number of these contours and thereby obtain the Chern number of the final Hamiltonian. 
With respect to the detuning $\delta$, we obtain the phase diagram shown in Fig.\,\ref{fig:8_PhaseDiagram}{\bf b}. It features a topologically non-trivial region with Chern number $1$ for a finite interval of detunings (corresponding to small values of $\Delta^\text{eff}$) surrounded by topological trivial regions. The measured Chern number agrees well with the theoretical prediction obtained from a numerical simulation (see methods).

{\bf Measurement of the micromotion}
In order to get a better resolution of the vortex dynamics, we measure the dynamics in steps of a quarter of the driving period $T=2\pi/\omega=156\,\mu$s. We thus sample the micromotion of the Floquet system \cite{Goldman2014,Bukov2015,Eckardt2015,Desbuquois2017}. Because the micromotion of the vortex positions is small compared to the contours of their trajectories in our case \cite{supmat}, it has no influence on the measurement of the Chern number. 

\begin{figure}[tb]
	\centering
		\includegraphics[width=0.8\linewidth]{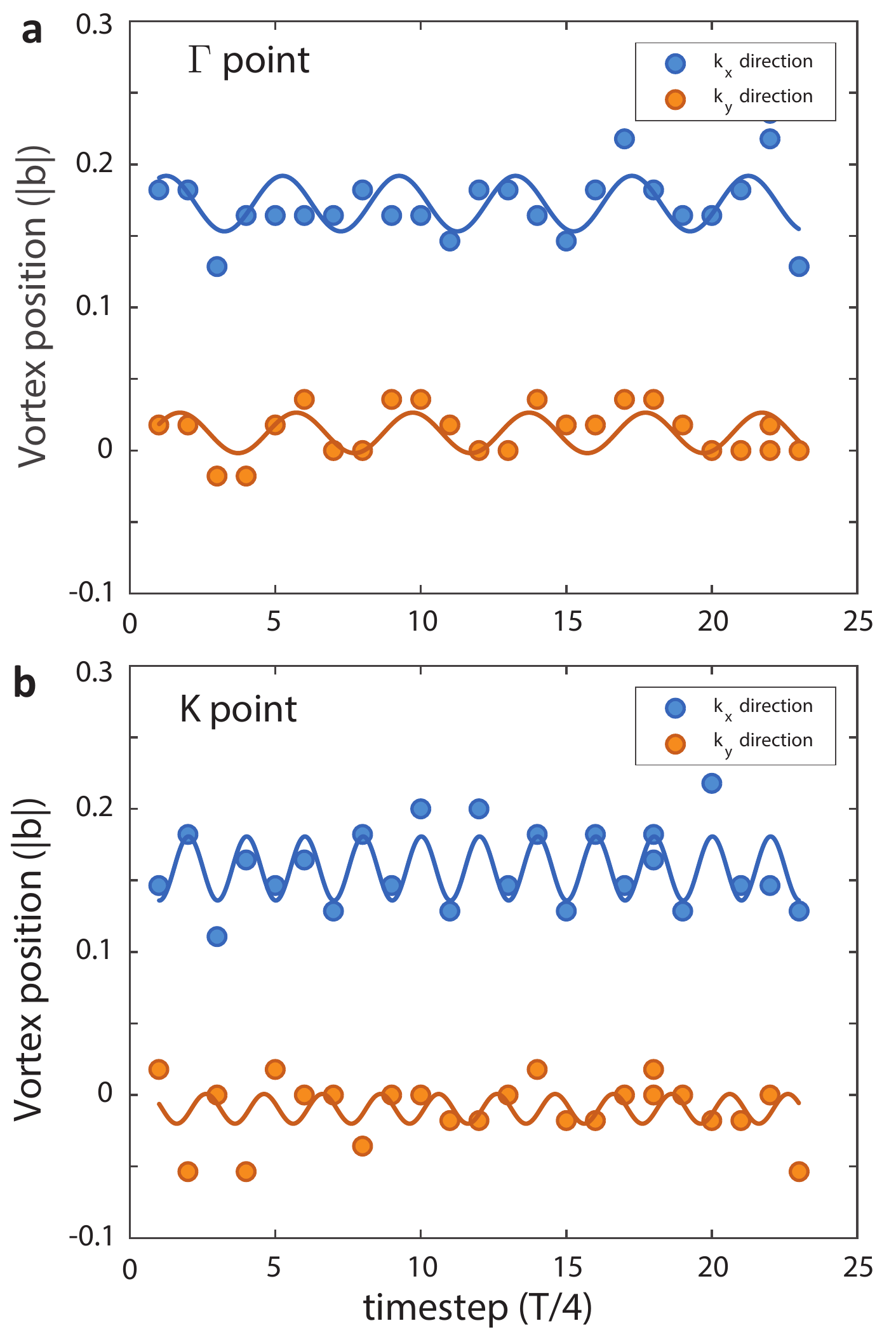}
		\caption{Micromotion of the static Dirac points. Position of the static vortices for the observed time steps (multiples of 39 $\mu$s, driving period $T$ is four time steps). {\bf a} Static vortex at the $\Gamma$ point. {\bf b} Static vortex at one of the $K$ points. The $k_x$ position is shifted by $+0.2|{\bf b}|$ in {\bf a} and $+0.7|{\bf b}|$ in {\bf b} for better visibility. The motion is approximately circular with the $k_x$ direction (blue points) being out of phase with the $k_y$ direction (orange points). The vortex at the $\Gamma$ point moves dominantly with the driving frequency. The vortex at the $K$ point moves dominantly at twice the driving frequency. The lines show fits as a guide to the eye with the periods fixed at $T$ and $T/2$, respectively. Both amplitudes are very small (few percent of the lattice vector length $|{\bf b}|$). The positions are determined to the precision of a single pixel of the images, which corresponds to $0.018|{\bf b}|$. The detuning is $\delta/2\pi=-478$\,Hz.}\label{fig:9_micromotion}
		\end{figure}

In Fig.\,\ref{fig:9_micromotion} we evaluate the micromotion of the static Dirac points in the experimental data. We find an approximately circular motion with the driving frequency or multiples of it as predicted by the derivation presented in \cite{supmat}. As expected from the scaling of the micromotion with $J_{AB}/(\hbar\omega)$ which is in the order of $\sim0.1$, the amplitude of the micromotion is very small (few percent of the lattice vector length $|{\bf b}|$) and does not affect the measurement of the topology of the system.

\begin{figure*}
		\includegraphics[width=0.9\textwidth]{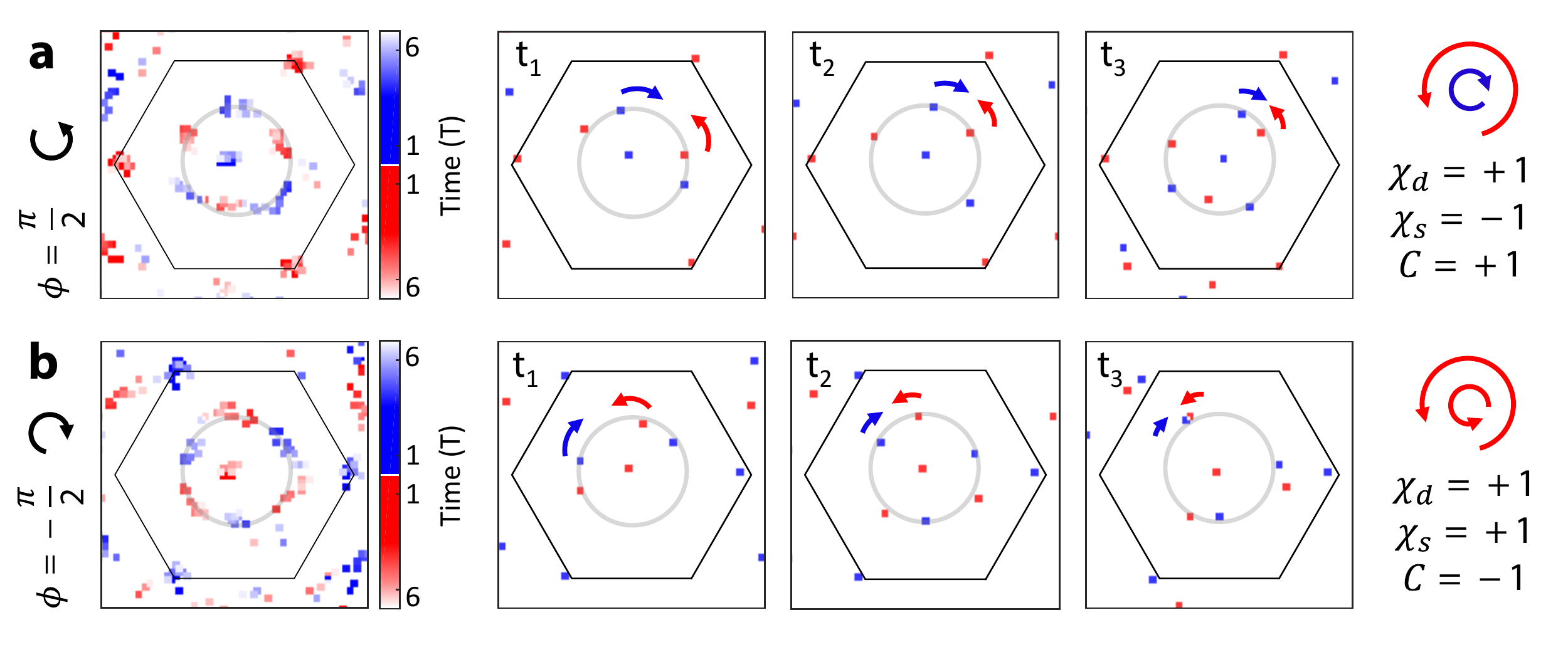}
		\caption{{\bf Sign of the linking number}. {\bf a}, Vortex data in the non-trivial regime (shaking phase of $\pi/2$ and shaking detuning of $\delta/2\pi=-372$\,Hz). The first subfigure shows the time-integrated data, while the other subfigures show successive stroboscopic time steps $t_1=13 \cdot T/4$, $t_2=17 \cdot T/4$, $t_3=21 \cdot T/4$ after the quench. The vortex contour has a positive chirality, while the enclosed static vortex has a negative chirality, revealing the Chern number $+1$ (see text). {\bf b}, Reverse shaking (grey point in Fig.\,\ref{fig:3_Haldane}) for $\delta/2\pi=-359$\,Hz and for time steps $t_1=14 \cdot T/4$, $t_2=18 \cdot T/4$, $t_3=22 \cdot T/4$ after the quench. The chirality of the enclosed vortex is now inverted and the Chern number is $-1$.
	}\label{fig:10_Sign}
	\end{figure*}

{\bf Extracting the sign of the Chern number.}
The sign of the Chern number can be obtained from an analysis of the chiralities of the observed vortices and their direction of motion (see Fig.\,\ref{fig:10_Sign}) (compare ref.~\cite{Wang2017}). We can define a chirality $\chi_{\rm d}$ of the dynamical vortex contour from the direction in which the dynamical vortices of positive chirality move. The sign of the linking number can then be defined as the product $-\chi_{\rm d}\chi_{\rm s}$ with $\chi_{\rm s}$ denoting the chirality of the enclosed static vortex. From this sign, one directly obtains the sign of the Chern number of the lower Bloch band \cite{supmat}. Fig.\,\ref{fig:10_Sign} shows time-resolved vortex data for two different directions of the circular lattice shaking, which leads to Chern numbers of opposite sign. While the chirality of the vortex contour is the same in both cases, the chirality of the enclosed vortex changes with the driving direction, directly indicating the opposite sign of the Chern number.

{\bf Discussion and outlook.}
In summary, we found experimental evidence that the Chern number, which characterizes topologically non-trivial properties of insulating equilibrium states of a quantum system, determines also properties of its dynamics far away from equilibrium. Namely, we observed that it directly corresponds to the linking number of the trajectories of ${\bf k}$-space vortices that emerge after a strong quench. Furthermore, we measured the instantaneous Chern number of the time-evolved state and found that it indeed remains zero under the unitary dynamics. We also identified the sign of the linking number from the chiralities of the vortices and their direction of motion. We show that state tomography yields the correct topological properties also for measurements in dispersive bands, which allows for broader non-trivial regions.

It is an interesting question in how far such a correspondence between topological properties in equilibrium and far from equilibrium can be generalized to other topological indices, such as, e.g., the W3 winding number characterizing Floquet topological phases \cite{Rudner2013,Quelle2017}, or to strongly interacting systems. Our experiments present a direct measurement and visualization of a topological index as opposed to the usual approach of infering the topology from the quantization of a response, e.g. the Hall conductance \cite{Aidelsburger2015} or circular dichroism \cite{Asteria2018}. 

\newpage

{\bf Methods}\\

{\bf System preparation}
The experiments start with an ultracold cloud of about $3\cdot 10^5$ spin-polarized $^{40}$K atoms in the $F=9/2$, $m_F=9/2$ state. We linearly ramp up the hexagonal optical lattice in 10 ms and hold for another 5 ms before switching on the lattice shaking. In the direction orthogonal to the hexagonal lattice, the sample is harmonically confined, i.e. realizing a lattice of tubes. The lattice is formed by the interference of three laser beams of wavelength $\lambda_L=1064$ nm and we introduce an $AB$-offset by polarization control of the beams\,\cite{Flaschner2016}. We image the sample on a CCD camera after 21 ms of time-of-flight expansion, which leads to a magnification where one lattice vector length $|\mbf{b}|$ corresponds to 56 pixel). The state tomography uses 32 time steps of 8 $\mu$s and a sinusoidal fit including an exponential damping (see Fig.\,\ref{fig:4_scheme}). 

{\bf Exact numerical simulation of the driven lattice}	
To obtain the tight-binding description of our lattice, we start from the known lattice geometry fixed by the polarization of the lattice beams (linear polarization tilted $9^{\mathrm{o}}$ out of the lattice plane, with a relative phase of the in-plane and out-of-plane polarization of 0, $2\pi/3$ and $4\pi/3$ for the three beams). We calculate the exact band structure for this geometry and different lattice depths $V_0$. To determine a precise value of the lattice depth, we use the band distance data of the bare lattice from the state tomography and fit the exact band structure to it \cite{supmat}. In this way, we compensate for small drifts of the lattice depth. We then fit a tight binding model to the exact band structure and obtain $\Delta_{AB}$, $J_{AB}$, $J_{AA}$ and $J_{BB}$. 
The values of $J_{AA}/h$ are in the range of 80 to 115\,Hz and $J_{BB}/h$ in the range of -2 to -6\,Hz. For the comparison with the effective Hamiltonian the small $J_{AA}$ and $J_{BB}$ are neglected.

We compare our data to exact numerics of the driven tight-binding model (Fig.\,\ref{fig:3_Haldane} and Fig.\,\ref{fig:8_PhaseDiagram}). In this calculation, the time evolution operator $U(t,0)$ is calculated via time slicing as a product of time evolution operators for constant Hamiltonians (compare ref.\,\cite{Flaschner2016}). This method works for any evolution time $t$ including sub-stroboscopic time steps, where the micromotion is automatically taken into account. It depends on the initial phase of the shaking, which we set to zero as in the experiment. To obtain a prediction for the dynamical vortex contours, we calculate the overlap of the time evolved state $|\psi(t)\rangle=U(t,0)|\psi(0)\rangle$ with the initial state $|\psi(0)\rangle$ and count momenta where this overlap is below a threshold of 0.02. In the last column of Fig.\,S3 of the Supplementary Material, these numerical data are summed up for all time steps (using a resolution of eight time steps per driving period, in order to better resolve the contours). The phase diagrams in Fig.\,\ref{fig:3_Haldane} and Fig.\,\ref{fig:8_PhaseDiagram} are obtained from this exact numerics and the Chern number is calculated as the integral of the Berry curvature. The non-integer values from the calculation on a finite grid in momentum space are removed by setting the Chern number to zero or one based on a threshold of 0.5.

\bibliographystyle{natbib.sty}

We acknowledge financial support from the Deutsche Forschungsgemeinschaft via the Research Unit FOR 2414 and the excellence cluster “The Hamburg Centre for Ultrafast Imaging - Structure, Dynamics and Control of Matter at the Atomic Scale”. BSR acknowledges financial support from the European Commission (Marie Curie Fellowship). We acknowledge fruitful discussions with Ramanjit Sohal and Christoph Sträter.
[Competing Interests] The authors declare that they have no competing financial interests.
[Correspondence] Correspondence and requests for materials should be addressed to K.S. \\(email: klaus.sengstock@physnet.uni-hamburg.de).



\setcounter{equation}{0}
\setcounter{figure}{0}
\setcounter{table}{0}
\makeatletter
\renewcommand{\theequation}{S\arabic{equation}}
\renewcommand{\thefigure}{S\arabic{figure}}
\renewcommand{\bibnumfmt}[1]{[S#1]}
\renewcommand{\citenumfont}[1]{S#1}


\newpage

\section{Supplemental material}

\appendix

\section{Tight-binding description}
We consider a system of spinless fermions in a hexagonal lattice with sublattice offset $\Delta = \nu\hbar\omega + \hbar\delta$ that is near-resonantly driven by a circular force $\bF(t) = -F[\cos(\omega t){\bm e}_x + \sin(\omega t){\bm e}_y]$ [See Fig.\,2(b)].  Here $\nu$ is an integer and $\hbar\delta\ll\hbar\omega$ the detuning. (Our experiment is described by $\nu=1$, whereas the case $\nu=0$ captures the Floquet topological insulator proposed in Ref.~\cite{OkaAoki09}, which was realized both with optical wave guides \cite{RechtsmanEtAl13} and in an optical lattice experiment \cite{JotzuEtAl14}). In general, the system is described by the Hubbard Hamiltonian
\be
\begin{aligned}
\Ho(t) = 
& -J\sum_{\la\ell'\ell\ra} \aa_{\ell'} \ao_{\ell} \\
         & + \sum_\ell\big[-\br_\ell\cdot\bF(t)
         + (\nu\hbar\omega + \hbar\delta) \delta_{\ell\in B}\big]\no_\ell,
\end{aligned}
\ee
where $\aa_\ell$, $\ao_\ell$, and $\no_\ell=\aa_\ell\ao_\ell$ denote the creation, annhilation, and number operator for fermions on lattice site $\ell$ at position $\br_\ell$, respectively, where $J$ describes tunneling between nearest neighbor pairs $\la\ell'\ell\ra$, and where $\delta_{\ell\in B}$ is one if $\ell$ lies in sublattice $B$ and zero otherwise. Since we do not consider any bare next-nearest-neighbor hopping, here we have dropped the subscript indices used in the main text ($J\equiv J_{AB}$ and $\Delta\equiv\Delta_{AB}$) in order to make the notation simpler. The force $\bF(t)$ is an inertial force created by moving the lattice along a circular orbit in space, so that the Hamiltonian describes the system in the reference frame co-moving with the lattice.

Let $|\psi(t)\ra$ denote the state of the system in the lattice frame. It is
convenient to perform a gauge transformation
$ |\psi'(t)\ra=\Uo^\dag(t)|\psi(t)\ra$
and $\Ho'(t) = \Uo^\dag(t)\Ho(t)\Uo(t)-i\hbar\Uo^\dag(t)\dot{\Uo}(t)$,
with the unitary operator
\be
\begin{aligned}
\Uo(t) & = \Uo_\text{shift}(t)\Uo_\text{rot}(t) \\
& = \exp\bigg(i\sum_\ell \big[\chi^{\text{shift}}_\ell(t)+\chi^{\text{rot}}_\ell(t)\big]\no_\ell\bigg),
\end{aligned}
\ee
where
\be
\begin{aligned}
\chi^\text{shift}_\ell(t) & = \frac{F}{\hbar\omega}\br_\ell\cdot
    [-\sin(\omega t){\bm e}_x + \cos(\omega t){\bm e}_y],\\
\qquad
\chi^\text{rot}_\ell(t) & = - \nu\omega t \delta_{ \ell \in B}.
\end{aligned}
\ee
While $\Uo_\text{shift}(t)$ integrates out the time-periodic shift in
quasimomentum induced by the circular force, $\Uo_\text{rot}(t)$ captures a
rotation of the pseudospin defined by the sublattice degree of freedom and
integrates out the resonant part $\nu\hbar\omega$ of the sublattice imbalance
$\Delta$. The resulting transformed Hamiltonian reads
\be
\Ho'(t) = -\sum_{\la\ell'\ell\ra} Je^{i\theta_{\ell'\ell}(t)}
                \aa_{\ell'} \ao_{\ell}
         + \sum_\ell \delta \delta_{\ell\in B} \no_\ell,
\ee
with time-periodic Peierls phases $\theta_{\ell'\ell}(t)
= \alpha \sin(\omega t -\varphi_{\ell'\ell}) - \sigma_{\ell}\nu \omega t$. Here, we have
introduced the dimensionless driving strength $\alpha = Fa/\hbar\omega$, with
$a=\frac{1}{\sqrt{3}}\frac{2}{3}\lambda_L$ denoting the distance between adjacent lattice sites, $\varphi_{\ell'\ell}$
denotes the azimuthal angle of the vector $\br_{\ell'}-\br_\ell$, and
$\sigma_{\ell}=1$ ($\sigma_{\ell}=-1$) for $\ell\in A$ ($\ell\in B$).

The transformation preserves the periodic time dependence of the Hamiltonian and
removes large energy offsets of order $\hbar\omega$ between neighboring sites.
With that it provides a good starting point for computing the effective
time-independent Hamiltonian $\Ho_F$ and the periodic micromotion operator
$\Uo_F(t)$ in a high frequency approximation \cite{EckardtAnisimovas15,
GoldmanDalibard14}, in terms of which the time-evolution operator for the dynamics induced by $\Ho^{\prime}(t)$ takes the transparent form
\be\label{eq:U'}
\Uo'(t,t_0) = \Uo_F(t)\exp\Big(-\frac{i}{\hbar}(t-t_0)\Ho_F\Big)\Uo^\dag_F(0).
\ee
Note that the transformation $\Uo(t)$ restores also the translational symmetry
of the lattice, which was broken by the on-site potential $-\br_\ell\cdot\bF(t)$,
so that the Floquet states of $\Ho'(t)$ and the eigenstates of $\Ho_F$ are Bloch
states.

\section{Effective Hamiltonian}
In order to compute the effective Hamiltonian, we will keep the two leading terms
of the high-frequency expansion \cite{EckardtAnisimovas15},
\be
\begin{aligned}
\Ho_F & \approx \Ho_F^{(1)}+\Ho_F^{(2)}
\quad\text{with}\quad
\Ho_F^{(1)} = \Ho_0, \\
\qquad 
\Ho_F^{(2)} & = \sum_{m=1}^\infty \frac{\big[\Ho_m,\Ho_{-m}\big]}{m\hbar\omega}.
\end{aligned}
\ee
Here,
\be
\begin{aligned}
\Ho_m & = \frac{1}{T} \int_0^T\! \rd t\, \Ho'(t) e^{-im\omega t} \\
       & = -\sum_{\la\ell'\ell\ra} J^{(m)}_{\ell'\ell} \aa_{\ell'} \ao_{\ell}
         + \delta_{m,0}\sum_\ell \hbar\delta \delta_{\ell\in B} \no_\ell
\end{aligned}
\ee
denote the Fourier components of the Hamiltonian, with tunneling parameters
$J^{(m)}_{\ell'\ell}
    = J \mathcal{J}_{m+\sigma_{\ell}\nu}(\alpha)
        e^{-i(m+\sigma_{\ell}\nu)\varphi_{\ell'\ell}}$, where
$\mathcal{J}_n(x)$ is an ordinary Bessel function of the first kind.

Evaluating these terms, we find
\be
\begin{aligned}
\Ho_F \approx
& - \sum_{\la\ell'\ell\ra} J^{\text{eff}}_{\la\ell'\ell\ra}\aa_{\ell'} \ao_{\ell}
 - \sum_{\lla\ell'\ell\rra} J^{\text{eff}}_{\lla\ell'\ell\rra}\aa_{\ell'}
        \ao_{\ell} \\
& +        \sum_\ell \Delta_\text{eff} \delta_{\ell\in B} \no_\ell , \label{eq H_F eff. floq. Ham.}
\end{aligned}
\ee
where $\lla\ell'\ell\rra$ denote pairs of next-nearest neighbors.
The effective nearest-neighbor tunneling matrix elements,
\be
J^{\text{eff}}_{\la\ell'\ell\ra}
    = J\mathcal{J}_{\sigma_{\ell}\nu}(\alpha) e^{-i\sigma_{\ell}\nu\varphi_{\ell'\ell}},
\ee
originate from the first-order term $\Ho_F^{(1)}$. In turn, the effective
next-nearest-neighbor tunneling matrix elements
\be
\begin{aligned}
J^{\text{eff}}_{\lla\ell'\ell\rra}
        = & - \sum_{m=1}^\infty \frac{J^2}{m\hbar\omega}
            \Big[\mathcal{J}^2_{m-\sigma_\ell\nu}(\alpha)
                    e^{i(m-\sigma_\ell\nu)(\pi+\sigma_{\lla\ell'\ell\rra}\pi/3)} \\
                & -\mathcal{J}^2_{m+\sigma_\ell\nu}(\alpha)
                    e^{i(m+\sigma_\ell\nu)(\pi+\sigma_{\lla\ell'\ell\rra}\pi/3)} \Big],  \label{eq J^eff_nnn hopping}
\end{aligned}
\ee
stem from the first-order term and can be understood as a superexchange process. Here, $\sigma_{\lla\ell'\ell\rra}=1$  ($\sigma_{\lla\ell'\ell\rra}=-1$) for tunneling clockwise (counterclockwise) around a hexagonal plaquette of the lattice. The effective sublattice offset
\be
\Delta_\text{eff} = \hbar\delta + \sum_{m=1}^\infty \frac{zJ^2}{m\hbar\omega}
                   \Big[\mathcal{J}^2_{m-\sigma_\ell\nu}(\alpha)
                         -  \mathcal{J}^2_{m+\sigma_\ell\nu}(\alpha) \Big]  ,
\ee
with coordination number $z=3$, possesses contributions from both orders.

\subsection{Comparison of models with and without initial sublattice offset}
There is a fundamental difference between the case $\nu\ne 0$; corresponding to our experiment with $\nu=1$, and the
experiments with the case $\nu=0$ described in Refs.~\cite{RechtsmanEtAl13, JotzuEtAl14}. For
$\nu=0$, nearest-neighbor tunneling is present already in the undriven system and
second-order next-nearest neighbor tunneling is a driving induced process.
Conversely, for $\nu\ne0$ nearest-neighbor tunneling has to be induced by the
driving (since it is off-resonant in the undriven lattice), while
second-order next-nearest-neighbor tunneling occurs already in the undriven
system. This fact is reflected in the behavior of the effective tunneling matrix
elements in the limit of small driving strength $\alpha$, where we have
\be
\begin{aligned}
J^{\text{eff}}_{\la\ell'\ell\ra} & \simeq J + \mathcal{O}(\alpha^2),
\qquad \\
J^{\text{eff}}_{\lla\ell'\ell\rra}
           &  \simeq \frac{\sqrt{3}\alpha^2}{4}\frac{J^2}{\hbar\omega}
                    e^{i\sigma_{\lla\ell'\ell\rra}\pi/2}
                    +\mathcal{O}(\alpha^4)
\quad\text{ for }\quad \nu=0,
\end{aligned}
\ee
whereas
\be
\begin{aligned}
J^{\text{eff}}_{\la\ell'\ell\ra}
   & \simeq \frac{\alpha}{2} J
    \sigma_\ell e^{-i\sigma_{\ell}\varphi_{\ell'\ell}}
   +\mathcal{O}(\alpha^3),\\
\qquad
J^{\text{eff}}_{\lla\ell'\ell\rra} & \simeq - \sigma_\ell \frac{J^2}{\hbar\omega}
    + \mathcal{O}(\alpha^2)
\quad\text{ for }\quad \nu=1.
\end{aligned}
\ee
Here, we have used that $\mathcal{J}_n(x)
=\frac{1}{|n|!}\big[\mathrm{sgn}(n)x/2\big]^{|n|}+\mathcal{O}(x^{|n|+2})$. The opposite sign of the effective next-nearest- neighbor tunneling on the two sublattices arises from the opposite sign of the offset to the intermediate state in the superexchange process. This difference between the cases $\nu=0$ and $\nu=1$ has two major consequences. The first one is related to the fact that the Peierls phases appear at the driving induced tunneling matrix elements. For $\nu=0$, the effective {next-nearest neighbor} tunneling matrix elements are complex, corresponding to the configuration of the Haldane model \cite{Haldane88}. In our case, for $\nu=1$, instead the \emph{nearest-neighbor} tunneling matrix elements acquire a phase. While the model can still be mapped to the Haldane model via a
gauge transformation, this implies that one of the Dirac cones is shifted from one of the $K$ points at the corner of the first Brillouin zone to the $\Gamma$ point at its center.
The second consequence is more important: The topologically non-trivial properties of the effective Hamiltonian emerge from the interplay between nearest-neighbor tunneling processes on the one hand and next-nearest-neighbors tunneling processes on the other. If the energy scale of one of these processes is much smaller than that of the other one, the topological band gap will be of the order of this smaller energy scale. For $\nu=0$ the next-nearest neighbor tunneling matrix elements, which are suppressed already by a factor of $J/(\hbar\omega)$ with respect to nearest-neighbor tunneling, scale only quadratically with the driving amplitude $\alpha$, so that for not too strong
driving the band gap scales like $\alpha^2 J^2/(\hbar\omega)$. In contrast, for $\nu=1$ the gap should roughly scale like $\alpha J$ as long as $\alpha\lesssim J/(\hbar\omega)$ and like $J^2/(\hbar\omega)$ for larger driving strength (as long as $\alpha\le 1$). This suggests that the case $\nu=1$ is favorable for the realization of robust topological band structures. Indeed, the width of the region with non-trivial Chern number is 100 Hz in Ref.~\cite{JotzuEtAl14}, but 500 Hz in this work. However, in an implementation with an inhomogeneous lattice, where the resonance condition of the global shaking varies across the sample, the ratio of the width of the non-trivial region to the driving frequency is also relevant.

\subsection{Effective Hamiltonian in quasimomentum representation}

It is instructive to express the effective Hamiltonian given in Eq.(\ref{eq H_F eff. floq. Ham.}) in quasimomentum representation,
\be
\Ho = \sum_\bk (\aa_{A\bk}, \aa_{B\bk} ) \,
      [h_0(\boldsymbol{k})\cdot \boldsymbol{I} + \boldsymbol{h}(\boldsymbol{k})\cdot \boldsymbol{\sigma}]
      \left({\ao_{A\bk}\atop\ao_{B\bk}}\right).
\ee
Here, $\boldsymbol{I}$ is $2\times2$ identity matrix, $\boldsymbol{\sigma}$ denotes the Pauli matrices acting on the pseudospin space defined by the two sublattice states $\sigma=A,B$, and $\ao_{\sigma\bk} =\frac{1}{\sqrt{M}}\sum_{\ell\in\sigma}e^{-i\bk\cdot\br_\ell}\ao_\ell$ the annihilation operator for a fermion with quasimomentum $\bk$ on sublattice $\sigma$, where $M$ is the number of lattice cells. The components of the Hamiltonian on the Bloch sphere follow as
\begin{align}  \label{eq_components}
&h_x^f(\boldsymbol{k})= -J \mathcal{J}_{\nu}(\alpha) \sum_{j=1}^3 \cos(\boldsymbol{k}\cdot\boldsymbol{a}_j-\nu \varphi_j),\qquad \\
&h_y^f(\boldsymbol{k})=  J \mathcal{J}_{\nu}(\alpha) \sum_{j=1}^3 \sin(\boldsymbol{k}\cdot\boldsymbol{a}_j-\nu \varphi_j),\qquad\\
&h_z^f(\boldsymbol{k})= -\frac{J^2}{\hbar\omega} c_z(\alpha) \sum_{j=1}^3  2\cos(\boldsymbol{k}\cdot\boldsymbol{b}_j) -\Delta_{\text{eff}}/2, \quad \\
&h_0^f(\boldsymbol{k})= -\frac{J^2}{\hbar\omega} c_0(\alpha) \sum_{j=1}^3  2\sin(\boldsymbol{k}\cdot\boldsymbol{b}_j) +\Delta_{\text{eff}}/2, \quad
\end{align}
where $\boldsymbol{a}_j$ is the vector that connects the nearest-neighbor sites, with $j$ labeling the three possible directions for moving from an $A$ site to a $B$ site, ${\bm a}_j=a[\cos(\varphi_j){\bm e}_x+\sin(\varphi_j){\bm e}_y]$ with their corresponding angles $\varphi_j$ defined from the positive $x$-axis. $\boldsymbol{b}_j$ denotes the lattice vectors ${\bm b}_1=a(\sqrt{3},0), {\bm b}_2=a(-\frac{\sqrt{3}}{2},\frac{3}{2}), {\bm b}_3=-{\bm b}_1-{\bm b}_2 $ which connect next-nearest neighbors, and $c_{0,z}(\alpha)$ are some constants coming from taking the sum in Eq.~(\ref{eq J^eff_nnn hopping}).

	\begin{figure*}[tb]
	\centering
		\includegraphics[width=\textwidth]{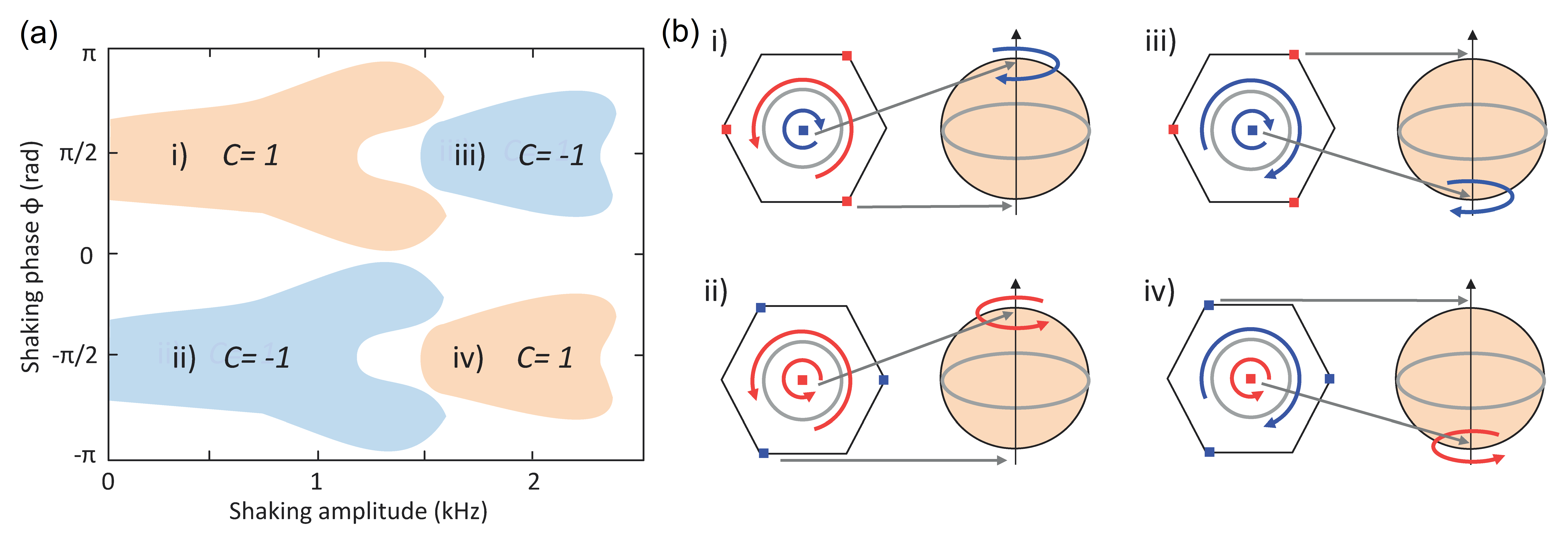}
		\caption{Sign of the Chern number. (a) In our Floquet system, the sign of the Chern number can change either by inverting the shaking direction or by increasing the shaking amplitude. Phase diagram calculated for a detuning of $\delta/2\pi=-40$\,Hz. (b) Sense of wrapping the Bloch sphere in the four different cases.}\label{fig:sign}
		\end{figure*}

\section{Micromotion}

In order to describe the influence of the periodic micromotion described by $\Uo_F(t)$ on the dynamics, let us
consider the first non-trival term of the high-frequency expansion
$\Uo_F(t)=\exp\big[\Go_1(t)+\Go_2(t)+\cdots\big]$. We approximate \cite{EckardtAnisimovas15}
\be
\Uo_F(t) \simeq \exp[\Go_1(t)],
\qquad
\Go_1(t)  =-\sum_{m\ne 0}^\infty\frac{\Ho_m e^{im\omega t}}
                {m\hbar\omega},
\ee
and find
\be
\begin{aligned}
\Go_1(t) & = \sum_{\la\ell'\ell\ra} g_{\la\ell'\ell\ra}(t) \aa_{\ell'}\ao_\ell,
\qquad \\
g_{\la\ell'\ell\ra}(t)
& = -\sum_{m\ne0}^\infty \frac{Je^{im\omega t}}{m\hbar\omega}
         \mathcal{J}_{m+\sigma_\ell\nu}(\alpha)
                    e^{-i(m+\sigma_\ell\nu)\varphi_{\ell'\ell}}.
\end{aligned}
\ee
This correction is of the same origin as the effective next-nearest neighbor
tunneling terms in the effective Hamiltonian. In leading order with respect to
the driving amplitude, the coefficients read
$g_{\la\ell'\ell\ra}(t)=-\alpha\frac{J}{\hbar\omega}
\cos(\omega t-\varphi_{\ell'\ell})$ for $\nu=0$ and $g_{\la\ell'\ell\ra}(t)
= \frac{J}{\hbar\omega}\sigma_\ell e^{-i\sigma_\ell\omega t}$ for $\nu=1$. For
$\nu=1$, this correction is again present already for infinitely weak driving, i.e. for $\alpha\rightarrow0$.

The operator $\Go_1(t)$ describes a time-periodic micromotion in real space,
where a particle at a given site $\ell$ explores neighboring lattice
sites. With respect to quasimomentum, it can be expressed like
\be
\Go_1(t) = \sum_\bk (\aa_{A\bk}, \aa_{B\bk} ) \,
      [g_x(\bk,t)\sigma_x+g_y(\bk,t)\sigma_y]\left({\ao_{A\bk}\atop\ao_{B\bk}}\right).
\ee
Here,
$g_x(\bk,t) = \mathrm{Re}\big(g(\bk,t)\big)$ and
$g_y(\bk,t) = \mathrm{Im}\big(g(\bk,t)\big)$, with
$ g(\bk,t) = -i\sum_{m\ne0}\sum_{j=1}^3 g_j(t) e^{-{\bm a}_j\cdot\bk}$,
where $g_j(t)$ describes $g_{\la\ell'\ell\ra}(t)$ for processes connecting an $A$
site $\ell$ with a neighboring $B$ site $\ell'$ at
$\br_{\ell'}=\br_\ell+{\bm a}_j$.
For $\nu=1$, we find
\be
\begin{aligned}
&g(\bk,t)
     = \sum_{m\ne 0}\frac{J}{m\hbar\omega}\mathcal{J}_{m+1}(\alpha) \times \\ 
     &\sum_{j=1}^3
    \exp\Big(m\omega t -{\bm a}_j\cdot\bk- (m+1)\varphi_j+\pi/2\Big).
\end{aligned}
\ee
For small driving amplitudes $\alpha$, the leading contribution stems from the
$m=-1$ term. Neglecting all other terms, $\Uo_F(t)$ describes a rotation in
pseudospin, by a $\bk$-dependent angle $\sim J/(\hbar\omega)$ around an axis in
the $xy$-plane that itself rotates around the z-axis with angular velocity
$\omega$ and $\bk$-dependent phase. Increasing $\alpha$, however, for
$\alpha\simeq 1$ both the the $m=-2$ term and the $m=1$ term become relevant so
that also higher harmonics of the driving frequency will make themselves felt in
the micromotion described by $\Uo_F(t)$.

Apart from the real-space micromotion described by $\Uo_F(t)$, another contribution to the micromotion is given by the transformation
$\Uo(t)=\Uo_\text{shift}(t)\Uo_\text{rot}(t)$ back to the original lattice frame of reference. It describes a phase rotation between different lattice sites, which corresponds to both a shift in quasimomentum and a rotation around the z-axis of the sublattice pseudospin.
Moreover, there is another effect. In order to predict the dynamics observed in the experiment, we also have to consider the experimental protocol, where lattice shaking is switched on at time $t_0$ and switched off again at the measurement time $t$. The shaking is performed in such a way that the relative lattice position $\bx(t+t^{\prime})$ changes continuously when the shaking is switched on. It is given by $\bx(t+t^{\prime})=\bz$ for $t+t^{\prime}<t_0$, $\bx(t+t^{\prime})=\bxi(t+t^{\prime})-\bxi(t_0)$ for $t_0<t+t^{\prime}<t$,
and by $\bx(t+t^{\prime})=\bxi(t)-\bxi(t_0)$ for $t+t^{\prime}>t$, where
$\bxi(t+t^{\prime})= -\Delta x [\cos(\omega (t+t^{\prime})){\bm e}_x+\sin(\omega (t+t^{\prime})){\bm e}_x]$ with
$\Delta x = F/(M\omega^2)$ and atomic mass $M$. Accordingly the lattice
velocity $\dot\bx(t+t^{\prime})$ is discontinuous, featuring jumps by $\dot\bxi(t_0)$ and
$-\dot\bxi(t))$ at $t+t^{\prime}=t_0$ and $t+t^{\prime}=t$, respectively. As a result the inertial
force $\bF_\text{ini}(t+t^{\prime})=-M\ddot\bx(t+t^{\prime})$ induced in the lattice frame of
reference, which is given by $\bF(t+t^{\prime})=-M\ddot\bxi(t+t^{\prime})$ between $t_0$ and $t$ and
vanishes before and after that, possesses also a contribution
$\bF_\text{boost}(t+t^{\prime})=-M\dot\xi(t_0)\delta(t+t^{\prime}-t_0) -\dot\xi(t)\delta(t-t)$.
These boosts shift the system's state in quasimomentum by $\bq(t_0)$ and
$-\bq(t)$ with $\bq(t+t^{\prime})=-(M/\hbar)\dot\bxi(t+t^{\prime})$, which is described by the
unitary operator $\Uo_{\bq}=\exp(i\sum_\ell \bq\cdot\br_\ell \no_\ell)$.
Thus, starting from the trivial insulator state $|\psi_0\ra$ at time $t+t^{\prime}<t_0$,
for times $t^{\prime}>0$ the time-evolved state reads
\be
|\psi(t+t^{\prime})\ra
=e^{-\frac{i}{\hbar}t^{\prime}\Ho^t}|\psi(t)\ra,
\ee
where $\Ho^t$ denotes the static tomography Hamiltonian describing the system
for times $t^{\prime}>0$ and where the state to be measured is given by
$ |\psi(t)\ra =
\Uo_{-\bq(t+t^{\prime})}\Uo_\text{shift}(t)\Uo_\text{rot}(t)
\Uo'(t,t_0) 
\Uo_\text{rot}^\dag (t_0)\Uo_\text{shift}^\dag(t_0)\Uo_{\bq(t_0)}|\psi_0\ra$.
Employing Eq.~(\ref{eq:U'}) as well as the fact that
$\Uo_\text{shift}(t+t^{\prime})=\Uo_{\bq}(t+t^{\prime})$, we find
\be
|\psi(t)\ra=
\Uo_\text{rot}(t)\Uo_F(t)e^{-\frac{i}{\hbar}(t-t_0)\Ho_F}
\underbrace{\Uo^\dag_F(t_0)\Uo_\text{rot}^\dag (t_0)|\psi_0\ra}
    _{\equiv|\psi'_0\ra} . \label{eq psi_micromotion}
\ee
Thus, the full micromotion, as it can be observed in the experiment is described
by
\be
\Uo_\text{micro}(t)=\Uo_\text{rot}(t)\Uo_F(t) .
\ee
One should note that $|\psi_0^{\prime}\ra=\Uo_\text{micro}^\dag (t_0) |\psi_0\ra$ is not an eigenstate of the initial Hamiltonian. We can overcome this by transforming the initial Hamiltonian as well; $\Ho^{i\prime}=\Uo_\text{micro}^\dag (t_0) \Ho^{i} \Uo_\text{micro} (t_0)$. When we also rotate the tomography Hamiltonian $\Ho^{t\prime}=\Uo_\text{micro}^\dag (t) \Ho^{i} \Uo_\text{micro} (t)$, it is now clear that these tomography and initial Hamiltonians are equal to each other only for tomography times $t=t_0+nT$ with integer $n$. For any other sub-stroboscopic time steps, the tomography Hamiltonian will not be parallel to the initial Hamiltonian on the Bloch sphere. In Fig.\,4(a) of the main text, we omit these contributions due to the micromotion and just aim to illustrate the experimental procedure.

The Hamiltonian $\Ho^t$ is represented by a quasimomentum-dependent vector
$\bh^t(\bk)$ playing the role of a magnetic field with respect to the sublattice
pseudospin and the state $|\psi(t)\ra$ is represented by a
quasimomentum-dependent unit vector $\hat{\psi}(t)$ denoting a point on the
Bloch sphere of that pseudospin. The positions of the measured vortices
correspond to those points in $\bk$-space, where both vectors are parallel (or
antiparallel). Thus, as long as $\bh^t(\bk)$ points to the south (or north) pole
everywhere, the pseudospin rotation $\Uo_\text{rot}(t)$ at angular velocity
$\omega$ will not make itself felt. However, as soon as $\bh^t(\bk)$ tilts away
from the north pole, as it is the case in the present experiment, this rotation
will cause an oscillatory behavior of the vortex position with respect to the
time $t$. Thus, the interplay between the oscillations induced by
$\Uo_F(t)$ and that by $\Uo_\text{rot}(t)$ is another source for the
generation of higher harmonics in the motion of the vortex position observed in
the experiment.

\section{Sign of the linking number}

We determine the sign of the linking number by comparing the relative chirality of the static ($\chi_s$) and dynamic vortices ($\chi_d$), i.e. the total sign is set by $\chi_s\chi_{d}$. The chirality of the dynamic vortex contour is given by the multiplication of the chirality of a vortex (or an antivortex) $\chi_{v}$ and the chirality of the path that it travels $\chi_{p}$. Since the dynamic vortex contour is the inverse image of the equator of the Bloch sphere, the direction of the motion is set by the gradient of the Hamiltonian $|\boldsymbol{h}(\boldsymbol{k}_v)|$ at the equator. This direction can be reversed by modifying the magnitude of the gap parameter $|\boldsymbol{h}(\boldsymbol{k_v})|$, without closing the gap at the Dirac point itself, i.e. without changing the chirality of the static vortex at the Dirac point. In the following, we show that changing the gradient of the Hamiltonian at the equator also converts the vortices into antivortices, hence, preserves the chirality $\chi_d$ and with that also the sign of the linking number. This means that the chirality of the dynamic vortex contour $\chi_d$ reflects indeed the topology of the Hamiltonian and cannot be changed by topologically-trivial deformations of the energy band. Our definitions are inspired by a related argument in ref. \cite{Wang2017Sup}.

The state of the system $\boldsymbol{\psi}(\boldsymbol{k},t)$ is given by Eq.~(\ref{eq w.f.}) for $\theta({\boldsymbol{k}},t)$ and $\phi({\boldsymbol{k}},t)$. The initial state points to the south pole for all $\boldsymbol{k}$, $\boldsymbol{\psi}(\boldsymbol{k},0)=-\hat{e}_z$ (in the case of dispersive bands, after performing the rotation given in Eq.~(\ref{eq rotation})). Quenching to the Floquet Hamiltonian induces a rotation by the angle $\alpha(\boldsymbol{k},t)\equiv\omega(\boldsymbol{k})t=2\boldsymbol{h}^f(\boldsymbol{k})t$ around the direction of $\hat{\boldsymbol{h}}^f(\boldsymbol{k})=\boldsymbol{h}^f(\boldsymbol{k})/|\boldsymbol{h}^f(\boldsymbol{k})|$. The time evolved state thus reads $\boldsymbol{\psi}(\boldsymbol{k},t)=\boldsymbol{R}(\alpha(\boldsymbol{k},t),\hat{\boldsymbol{h}}^f(\boldsymbol{k})) \boldsymbol{\psi}(\boldsymbol{k},0)$ where $\boldsymbol{R}(\alpha(\boldsymbol{k},t),\hat{\boldsymbol{h}}^f(\boldsymbol{k}))$ denotes the rotation matrix. For the given initial conditions, this gives
\begin{equation}\label{eq. time evolv. w.f.}
\boldsymbol{\psi}(\boldsymbol{k},t)=
\begin{pmatrix}
\hat{\boldsymbol{h}}_x \hat{\boldsymbol{h}}_z [1-\cos(\alpha(\boldsymbol{k},t))]+\hat{\boldsymbol{h}}_y \sin(\alpha(\boldsymbol{k},t)) \\
\hat{\boldsymbol{h}}_y \hat{\boldsymbol{h}}_z [1-\cos(\alpha(\boldsymbol{k},t))]-\hat{\boldsymbol{h}}_x \sin(\alpha(\boldsymbol{k},t))\\
\hat{\boldsymbol{h}}^2_z [1-\cos(\alpha(\boldsymbol{k},t))]+ \cos(\alpha(\boldsymbol{k},t))
\end{pmatrix}.
\end{equation}

In the tomography, we observe a static vortex whenever $\hat{\boldsymbol{h}}^f(\boldsymbol{k})\parallel \hat{e}_z$. Dynamic vortices occur when $\hat{\boldsymbol{h}}^f(\boldsymbol{k})\perp \hat{e}_z$, i.e. when $\hat{\boldsymbol{h}}^f(\boldsymbol{k})$ lies on the equator, so that for $\alpha(\boldsymbol{k},t_n)=n\pi$ and $\boldsymbol{\psi}(\boldsymbol{k},t_n)$ points to the north (south) pole for odd (even) integers $n$, where $t_n=n\pi/2|\boldsymbol{h}^f(\boldsymbol{k})|$. In the following, we will focus on the case $n=1$, where a dynamic vortex is found at time $t_1(\boldsymbol{k})$. The condition $\alpha(\boldsymbol{k},t_n)=\pi$ defines the trajectories of the dynamic vortices in quasimomentum $\boldsymbol{k}$, corresponding to the inverse image, $\mathbb{P}$, of the equator of the Bloch sphere with respect to the map $\hat{\boldsymbol{h}}(\boldsymbol{k}): \: \boldsymbol{k} \rightarrow \hat{\boldsymbol{h}}$. Note that in Ref.~\cite{Wang2017Sup}, this corresponds to the inverse image of the north pole with respect to the map $[\boldsymbol{k},t]\rightarrow \hat{\boldsymbol{h}}$.

\subsubsection{Direction of vortex motion}

Let $\boldsymbol{k}_v \,\epsilon\, \ell \subset\mathbb{P} $ be a point on the line $\ell$ which lies in the inverse image of the equator and $\hat{e}_{\parallel}(\boldsymbol{k}_v)$ denote a tangential unit vector of $\ell$ at $\boldsymbol{k}_{v}$ which defines a direction on this line. Then, the vortex which passes $\boldsymbol{k}_v$ at time $t(\boldsymbol{k}_v)$ moves with velocity $\boldsymbol{\dot{k}}_v=\dot{k}_{\parallel}\hat{e}_{\parallel}$ where $\dot{k}_{\parallel}=-\omega(\boldsymbol{k}_v)/\boldsymbol{g}_{\parallel}(\boldsymbol{k}_v)$ with $\boldsymbol{g}_{\parallel}(\boldsymbol{k})=t(\boldsymbol{k})\boldsymbol{\nabla}_{\boldsymbol{k}}\omega(\boldsymbol{k})\cdot \hat{e}_{\parallel}(\boldsymbol{k}_v)$.
Thus, as long as the gap does not close $[\omega(\boldsymbol{k})>0]$, the direction of motion $\chi_p$ is determined by the gradient of the gap along the line $\ell$,
\begin{equation} \label{eq. chi_p}
\chi_p=-\text{sgn}[\boldsymbol{g}_{\parallel}(\boldsymbol{k}_v)].
\end{equation}
Here, $\hat{e}_{\parallel}$ is the unit vector obtained from $\boldsymbol{\nabla}_{\boldsymbol{k}}\hat{h}_z(\boldsymbol{k})$ by an azimuthal rotation by $\pi/2$; $\hat{e}_{\parallel}=\boldsymbol{R}(\hat{e}_{z},\pi/2) \boldsymbol{\nabla}_{\boldsymbol{k}}\hat{h}_z(\boldsymbol{k})$.

\subsubsection{Vortex chirality}
In order to obtain the chirality of a dynamic vortex at point $\boldsymbol{k}_v$, we expand the wave function $\psi(\boldsymbol{k}_v,t(\boldsymbol{k}_v))$ in the vicinity of $\boldsymbol{k}_v$,
\begin{equation}\label{eq w.f. taylor}
\boldsymbol{\psi}(\boldsymbol{k})=
\begin{pmatrix}
0 \\
0\\
-1
\end{pmatrix} +\delta\boldsymbol{k}
\begin{pmatrix}
\hat{\boldsymbol{h}}_x(\boldsymbol{k}_v) \boldsymbol{f}(\boldsymbol{k}_v) - \hat{\boldsymbol{h}}_y(\boldsymbol{k}_v) \boldsymbol{g}(\boldsymbol{k}_v) \\
\hat{\boldsymbol{h}}_y(\boldsymbol{k}_v) \boldsymbol{f}(\boldsymbol{k}_v) - \hat{\boldsymbol{h}}_x(\boldsymbol{k}_v) \boldsymbol{g}(\boldsymbol{k}_v) \\
0
\end{pmatrix},
\end{equation}
where $\boldsymbol{f}(\boldsymbol{k}_v)=2\boldsymbol{\nabla}_{\boldsymbol{k}}\hat{h}_z(\boldsymbol{k})$ and $\boldsymbol{g}(\boldsymbol{k}_v)=\pi/\boldsymbol{h}(\boldsymbol{k}) \boldsymbol{\nabla}_{\boldsymbol{k}} |\boldsymbol{h}(\boldsymbol{k})|$ as before. This can be also expressed as
$ \delta\boldsymbol{\psi}(\boldsymbol{k})= \boldsymbol{f}(\boldsymbol{k}_v)\delta\boldsymbol{k} \hat{\boldsymbol{h}}(\boldsymbol{k}) +\boldsymbol{g}(\boldsymbol{k}_v)\delta\boldsymbol{k} \hat{\boldsymbol{h}}^{\prime}(\boldsymbol{k})$, where
$ \hat{\boldsymbol{h}}^{\prime}(\boldsymbol{k})= (-\hat{\boldsymbol{h}}_y(\boldsymbol{k}_v), \hat{\boldsymbol{h}}_x(\boldsymbol{k}_v)) $
is a unit vector orthogonal to $\hat{\boldsymbol{h}}(\boldsymbol{k}_v)$ and that, like $\hat{\boldsymbol{h}}(\boldsymbol{k}_v)$, lies on the equator. These two unit vectors ($\hat{\boldsymbol{h}},\,\hat{\boldsymbol{h}}^{\prime}$) span a coordinate system that is rotated by $\phi(\boldsymbol{k}_v)$ with respect to the one spanned by ($\hat{e}_x,\,\hat{e}_y$).

The chirality $\chi_v$ of a dynamical vortex is now determined by whether the azimuthal phase $\phi(\boldsymbol{k})$ winds in positive or negative direction while $\delta\boldsymbol{k}$ is taken around a closed loop; $\delta\boldsymbol{k}=\delta k [\cos(\gamma)\hat{e}_x+\sin(\gamma)\hat{e}_y]$ for $\gamma:0\rightarrow2\pi$. The chirality reads,
\begin{equation}\label{eq chi_d}
\chi_v=\text{sgn}[\boldsymbol{g}(\boldsymbol{k}_v)\times\boldsymbol{f}(\boldsymbol{k}_v)] =\text{sgn}[\boldsymbol{g}(\boldsymbol{k}_v)\cdot \hat{e}_{\parallel} ].
\end{equation}

Therefore, both the direction a vortex travels and its chirality depend on the gradient $\boldsymbol{g}(\boldsymbol{k}_v)$ of the gap at the vortex position $\boldsymbol{k}_v$. Inverting the direction of the motion requires to invert the gradient of the gap $\boldsymbol{g}_{\parallel}(\boldsymbol{k}_v)$ along the line $\ell$. On the other hand, inverting the chirality of the vortex via a change of $\boldsymbol{g}(\boldsymbol{k}_v)$ requires to invert $\boldsymbol{g}(\boldsymbol{k}_v)\cdot\boldsymbol{f}_{\perp}(\boldsymbol{k}_v)\equiv \boldsymbol{g}_{\boldsymbol{f}_{\perp}}(\boldsymbol{k}_v)$. Note that the unit vector $\hat{e}_{\parallel}$, which is defined to point along the direction where $\hat{\boldsymbol{h}}_z(\boldsymbol{k}_v)$ keeps the constant value zero, stands perpendicular to the gradient of $\hat{\boldsymbol{h}}_z(\boldsymbol{k}_v)$. Thus, changing the direction of motion of the vortex without closing the gap implies that the vortex changes its chirality, which preserves the overall sign of the dynamic vortex counter $\chi_{v}\chi_p$. Hence, any deformation in the Hamiltonian that does not change the topology cannot change the observed sign of the linking number.

\subsection{Sign of the Chern number in our system}
The sign of Chern number is given by the sense in which the Bloch sphere is covered. This is fixed by (i) which one of the two Dirac points is at the north pole of the Bloch sphere and (ii) in which sense the states wrap around the Bloch sphere azimuthally. These questions can be simply related to the chiralities of the observed static and dynamical vortices: the chirality of the dynamical vortex contour $\chi_d$ determines, which Dirac point is at the north pole, while the chirality of the enclosed static vortex $\chi_s$ determines the azimuthal winding of the states. For our choices of sign conventions, the sign of the Chern number of the lowest band is given by $\text{sgn}[C]=-\chi_d \chi_s$.

The four different possible combinations of $\chi_d=\pm 1$ and $\chi_s=\pm 1$ can be realized in our system by changing the shaking parameters: (i) the occupation of the poles by the Dirac points inverts for large shaking amplitudes and (ii) the sense of wrapping around azimuthally inverts with the direction of shaking. Data for the two shaking directions at small shaking amplitude is presented in Fig.\,10 of the main text. Fig.\,\ref{fig:sign} illustrates the four different possibilities in the phase diagram spanned by shaking phase and shaking amplitude.

From the analysis of $\chi_d$ and $\chi_s$, one can therefore also distinguish whether the Chern number changes sign due to a different shaking direction or due to a large driving amplitude, where the sign change of the Chern number originates from the sign change of the Bessel function renormalization of the tunnel elements. Such phases have been realized in cold atoms \cite{Flaschner2016Sup} and helical photonic waveguides \cite{Guglielmon2018Sup}.

	\begin{figure}[tb]
	\centering
		\includegraphics[width=0.9\linewidth]{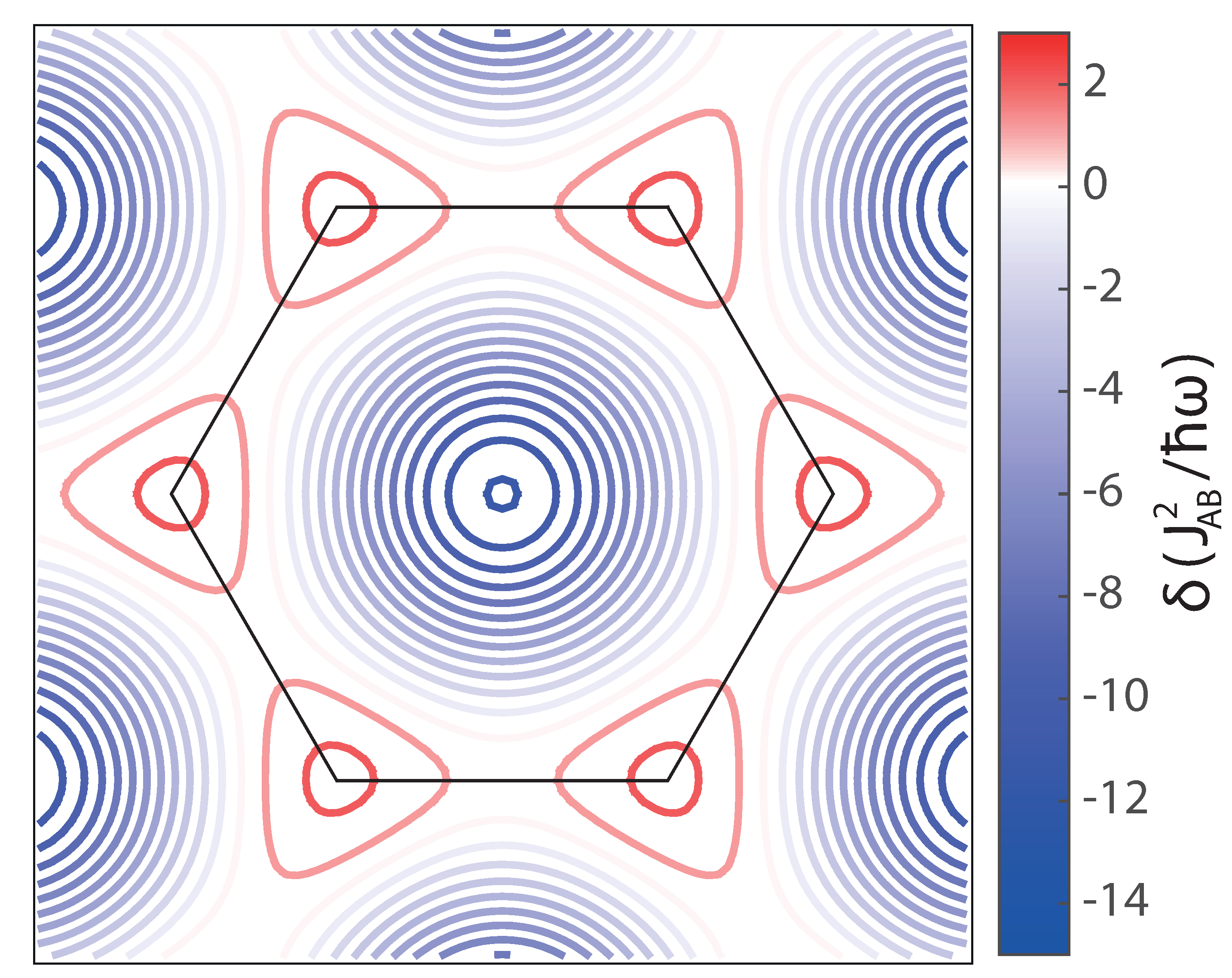}
		\caption{Dynamical vortex contours. Countors of the dynamical vortices for different shaking detunings according to the simple model described in the text.}\label{fig:contour}
		\end{figure}

\subsection{Experimental data for different detunings}
The experimental vortex data for different detunings leading to the phase diagram of Fig.\,8 is presented in Fig.\,\ref{fig:fulldata}. The different rows correspond to different detunings ranging from -938\,Hz to 515\,Hz. The detuning is varied by changing the lattice depth at fixed driving frequency of 6.410\,kHz and driving amplitude of 1\,kHz. The first column shows the band gap between the two lowest bare bands as obtained from the oscillation frequency of the tomography along three equivalent high symmetry paths (red, blue and green. The data is averaged over the six first time steps after the quench into the Floquet system). The collapse of the curves indicates the good balance of the three lattice beam intensities. The lattice parameters are obtained from a fit of the exact band structure to the band gap (black curve, regions around the static vortices are excluded from the fit). The extracted detuning and next-neighbor tunneling are states in the respective subfigures. The dashed horizontal line indicates the shaking frequency and indicates the near-resonant nature of the driving. 

The second column shows experimental data with the time-integrated static and dynamic vortices after the quench into the respective Floquet system. The hue indicates the time after the quench, at which the vortex appeared (lighter color means later time). The static vortices at the $\Gamma$ and $K$ points are present in all images. For non-trivial Chern number and in a regime of larger detunings, closed contours of dynamical vortices appear. The contour calculated from the effective Hamiltonian (green line) gives a reasonable approximation for the contour shape in the non-trivial regime. 

The third column shows the expected dynamical vortices from a full numerical calculation including the initial state and the micromotion. Each black dot indicated a zero scalar product between the initial state and the time-evolved state.

\subsection{Dynamical vortex contours from the effective Hamiltonian}

\begin{figure*}[!ht]
	\centering
		\includegraphics[width=0.7\textwidth]{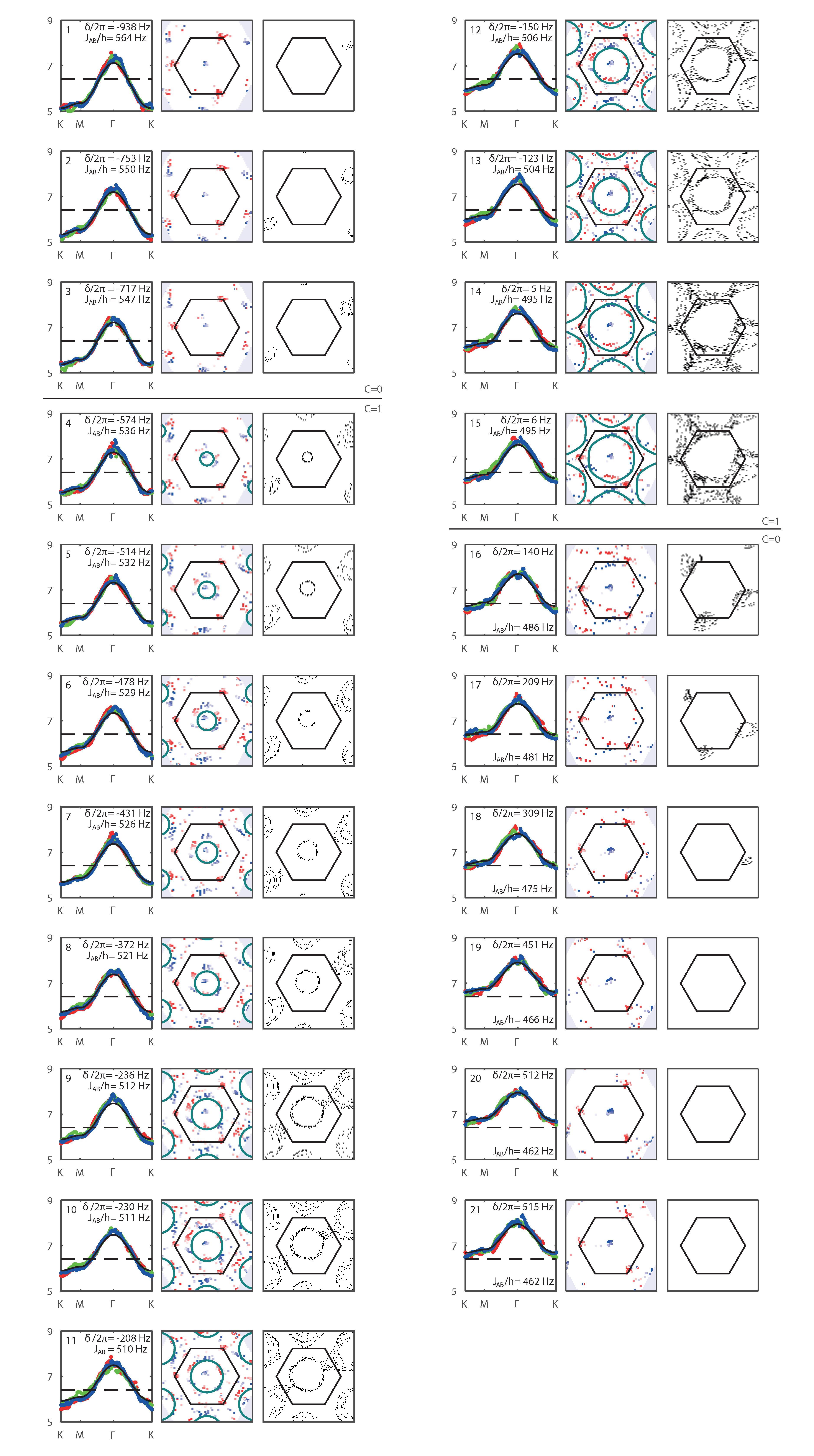}
		\caption{Experimental data for different detunings and comparison to calculations. See text for details.}\label{fig:fulldata}
		\end{figure*}

The effective Hamiltonian allows deriving a simple estimate for the dynamical vortex contours (green lines in Fig.\,\ref{fig:fulldata}). We neglect here the dispersion of the initial bands and the micromotion. In this approximation only non-trivial contours can be described, because the trivial contours arise from the finite dispersion of the initial bands. The vortex contour corresponds to the momenta, where the final Hamiltonian lies on the equator, i.e. where the $z$-component vanishes $h^{\mathrm{f}}_{z}(\mathbf{k})=0$. Using $h^{\mathrm{f}}_{z}(\mathbf{k})=\Delta^{\mathrm{eff}}/2+\sum_{j=1}^{3}(J_{AA}^{\mathrm{eff}}-J_{BB}^{\mathrm{eff}})\cos(\mathbf{k}\cdot \mathbf{b}_j)$ (compare Eq.\,\ref{eq_components}) and the effective tunneling elements in the low driving limit, this corresponds to
\begin{equation}
\begin{aligned}
S(\mathbf{k})& =\sum_{j=1}^{3}\cos(\mathbf{k}\cdot \mathbf{b}_j)=-\frac{\Delta^{\mathrm{eff}}}{2(J_{AA}^{\mathrm{eff}}-J_{BB}^{\mathrm{eff}})} \\
& =-\frac{\hbar\delta+3J_{AB}^2/\hbar\omega}{4J_{AB}^2/\hbar\omega}=-\frac{\tilde{\delta}+3}{4}
\end{aligned}
\end{equation}
with $\tilde{\delta}=\hbar\delta/(J_{AB}^2/\hbar\omega)$. The sum of the three cosines $S(\mathbf{k})$ can obtain values between $+3$ and $-3/2$. This means that contours only exist for detunings between $\tilde{\delta}=-15$ and $\tilde{\delta}=-+3$, which defines the non-trivial region.

Fig.\,\ref{fig:contour} shows the value of the sum of the three cosines $S(\mathbf{k})$, which correspond to the vortex contours for different detunings $\tilde{\delta}$. The contour closes around the $\Gamma$ point for $\tilde{\delta}=-15$ (where $S(\mathbf{k}=\Gamma)=3$) and around the K and K' points for $\tilde{\delta}=+3$ (where $S(\mathbf{k}=K)=-3/2$). In Fig.\,\ref{fig:fulldata} the contours are plotted together with the data for the respective detunings (green lines). While we don't expect quantitative agreement on this level of approximation, the predictions qualitatively explain the behavior of the data in the non-trivial regime. Furthermore, this discussion gives an intuitive picture for the detunings, where the topological phase transitions occur.

\bibliographystyle{mybibsty}

\end{document}